\documentclass[aps,superscriptaddress,nofootinbib,floatfix,epfs,preprintnumbers]{revtex4}
\usepackage{amsfonts,amscd,amsmath,amssymb,graphicx,color,float}
\usepackage[section]{placeins}
\def\ep{\text{e}}
\def\g{\mathsf{g}}
\def\oh{\frac{1}{2}}
\def\s{\mathsf{s}}
\def\m{\mathsf{m}}
\def\k{\mathsf{k}}
\def\n{\mathsf{n}}

\def\rq{r_q}
\def\rv{r_v}

\def\vv{v_{*}}
\def\Qqb{\text{\tiny Q}\bar{\text{\tiny q}}}

\def\QQb{\text{\tiny Q}\bar{\text{\tiny Q}}}
\def\Qqq{\text{\tiny Qqq}}
\def\QQq{\text{\tiny QQq}}
\def\QQ{\text{\tiny QQ}}
\def\3Q{3\text{\tiny Q}}
\def\qQb{\text{\tiny q}\bar{\text{\tiny Q}}}
\def\EI{E^{\text{\tiny (I)}}_{\QQq}}
\def\EII{E^{\text{\tiny (II)}}_{\QQq}}
\def\III{I^{\text{\tiny (II)}}}
\def\II{I^{\text{\tiny (I)}}}

\def\Qd{\text{\tiny [Qq]Q}}

\begin{document}
\preprint{LMU-ASC 33/20}
\title{Some Properties of the $QQq$-Quark Potential in String Models}
\author{Oleg Andreev}
 \affiliation{L.D. Landau Institute for Theoretical Physics, Kosygina 2, 119334 Moscow, Russia}
  \affiliation{V.A. Steklov Mathematical Institute, Gubkina 8, 119991, Moscow, Russia}
\affiliation{Arnold Sommerfeld Center for Theoretical Physics, LMU-M\"unchen, Theresienstrasse 37, 80333 M\"unchen, Germany}
\begin{abstract} 
We propose a string theory construction which allows us to study properties of the potential of two heavy quarks coupled to a light quark. In such a case, the potential is a function of separation between the heavy quarks. The results show the universality of the string tension and factorization at small separations expected from heavy quark-diquark symmetry. In addition, we make an estimate of the string breaking distance. With the parameter values we use, this distance is found to be almost the same as that for the heavy quark-antiquark potential. We also discuss the heavy quark-quark potential and its relation to Lipkin rule. 

\end{abstract}
\maketitle
\section{Introduction}
\renewcommand{\theequation}{1.\arabic{equation}}
\setcounter{equation}{0}

The success of the Cornell model \cite{cornell} in describing the quarkonium spectroscopy had a profound impact on the further development of potential models, as well as other methods of nonrelativistic QCD.\footnote{For more details and references, see the review articles \cite{rev}.} The key ingredient of those models is an assumption that, at leading order in the inverse heavy quark mass, the interaction between heavy quarks and antiquarks inside hadrons can be described by means of a potential. 

The recent discovery of the $\Xi_{cc}^{++}$ baryon through the LHCb experiments in CERN \cite{exp} has reinforced interest in the search for a theoretical description of double heavy baryons. The standard line of thought on a system composed by two heavy and one light quarks is that the heavy quarks may form compact bound states (diquarks) that allows one to effectively reduce the three-body problem to a two-body one. As a result, the properties of doubly heavy baryons turn out to be related to those of heavy mesons. This approach based on heavy quark-diquark symmetry \cite{wise} has been used for many years for making estimates in baryon spectroscopy \cite{bram}. 

Clearly, this symmetry is broken when the separation between the heavy quarks is not small enough. So, the question arises of whether another reduction to a two-body system can be achieved in this case. One way to answer this question is by the hadro-quarkonium picture \cite{voloshin} which is in fact an idea considered long ago in the context of interaction of charmonium inside nuclei \cite{stan}. Accordingly, the heavy quark pair is considered as being embedded in a light quark cloud. At leading order in the inverse heavy quark mass, the heavy quarks are static and the energy ($QQq$ potential) is determined by their relative separation and the representation of the underlying cylindrical symmetry group. Then one can use the resulting potential as an input to potential models, or to effective models of nonrelativistic QCD \cite{soto}. 

Although lattice gauge theory remains a basic tool for studying nonperturbative phenomena in QCD, it has had rather limited results on the $QQq$ potentials up to now \cite{suganuma,bali}. On the other hand, the gauge/string duality \cite{malda} provides new theoretical tools for studying strongly coupled gauge theories, and therefore may be used as an alternative method to treat this problem. 

In this paper, we propose a string theory construction for the $QQq$ system which allows us to study its ground state. Unlike lattice gauge theory, this issue has not been addressed in the context of the gauge/string duality, particularly in AdS/CFT. The paper continues a series of studies of the heavy quark potentials started in \cite{az1,a-hyb,a-3q}. The model we use to illustrate the basic ideas is not exactly dual to QCD, but it may be useful as a guide in the present period, when the string dual to QCD is unknown and the lattice results are still limited. In \cite{az1}, the quark-antiquark potential was derived. It is Coulomb-like at short distances and linear at long distances. Subsequent work \cite{white} made it clear that this model should be taken seriously, particular in the context of consistency with the available lattice data as well as phenomenology. In \cite{a-hyb}, the hybrid potential $\Sigma_u^-$ was modeled. The result is in a good agreement with the lattice. In addition, the result for the three quark potential, which smoothly interpolates between almost the $\Delta$-law as short distances and the $Y$-law at long distances, was presented in \cite{a-3q}. In this case the agreement with the lattice is also very good.  

In Sec.II of this paper, we review the relevant questions about the framework and particular model in which we will work. Then, in Sec.III, we construct and analyze a set of string configurations describing the $QQq$ system. The main result is to find the configurations which contribute to the ground state of the system. As a byproduct, we comment on a diquark $[Qq]$ appearing in one of the configurations. We also discuss the string breaking phenomenon and make an estimate of the string breaking distance. In Sec.IV we consider some properties of the heavy quark-quark potential. We show that the Lipkin rule does not hold in the model and the relation between the potentials is more complicated.\footnote{In the literature, the relation $E_{\QQ}=\oh E_{\QQb}$ between the heavy quark-quark and quark-antiquark potential is often called the Lipkin rule \cite{lipkin}, or alternatively $1/2$ rule \cite{jmr}.} We conclude in Sec.V with a discussion of some open problems. Additional technical details are included in the Appendix.

\section{Preliminaries}
\renewcommand{\theequation}{2.\arabic{equation}}
\setcounter{equation}{0}

In our discussion we will use a formalism proposed recently in \cite{a-strb1}. We illustrate most ideas with a simple model which purports to mimic QCD with two light flavors. So, we take the background to be 

\begin{equation}\label{metric}
ds^2=\ep^{\s r^2}\frac{R^2}{r^2}\Bigl(dt^2+d\vec x^2+dr^2\Bigr)
\,,
\qquad
{\text T}={\text T}(r)
\,,
\end{equation}
with $\text{T}$ a scalar field (open string tachyon). This geometry is one of the deformations of the Euclidean $\text{AdS}_5$ space of radius $R$, with a deformation parameter $\s$, which is used to mimic QCD. It is distinguished from the others primarily by its attractive features: computational simplification and phenomenological applications.\footnote{For some of the background, see \cite{az1,a-hyb,a-3q,white}.} The physical meaning of $\text T$ is that it describes light quarks at string endpoints in the interior of five-dimensional space that allows one to construct disconnected string configurations. This is one way to model the phenomenon of string breaking. We introduce a single scalar field (tachyon), since in what follows we consider only the case of two light quarks of equal mass.\footnote{The use of the term tachyon seems particularly appropriate in the context of instability of a QCD string.}

Just as for Feynman diagrams in field theory, we need some building blocks to construct string configurations. The first one is a Nambu-Goto string whose action is given 

\begin{equation}\label{NG}
S_{\text{\tiny NG}}=\frac{1}{2\pi\alpha'}\int d^2\xi\,\sqrt{\gamma^{(2)}}
\,,
\end{equation}
where $\gamma$ is an induced metric, $\alpha'$ is a string parameter, and $\xi^i$ are world-sheet coordinates. 

The next needed block is a baryon vertex. In AdS/CFT correspondence it is a five brane wrapped on an internal space \cite{witten}. At leading order in $\alpha'$, 
   the brane action is simply $S_{\text{vert}}={\cal T}_5\int d^6\xi\sqrt{\gamma^{(6)}}$, where ${\cal T}_5$ is a brane tension and $\xi^i$ are world-volume coordinates. Since the brane is wrapped on the internal space, it looks point-like in five dimensions. We assume that for the problem at hand the same holds if all objects are placed at the same fixed point in the internal space $\mathbf{X}$. If so, then what matters is a possible warp factor for $\mathbf{X}$. In \cite{a-3q}, it was observed  that an overall warp factor $\ep^{-\s r^2}$ yields very satisfactory results, when compared to the lattice calculations of the three quark potential. As usual we pick a static gauge $\xi^0=t$ and $\xi^a=\theta^a$, with $\theta^a$ coordinates on $\mathbf{X}$. The action is then 

\begin{equation}\label{baryon-v}
S_{\text{vert}}=\tau_v\int dt \,\frac{\ep^{-2\s r^2}}{r}
\,.
\end{equation}
Unlike in AdS/CFT, where $\tau_v={\cal T}_5R\,\text{vol}(\mathbf{X})$, here $\tau_v$ is treated as a free parameter. The reason for this is that the action provides an effective description of the five brane. It includes neither $\alpha'$-corrections to the leading term \footnote{For the value of $\g=0.176$ obtained from fitting of the lattice data, a simple estimate gives $\alpha'/R^2=1/2\pi\g\approx 0.9$.} nor Ramond-Ramond forms. 

Finally the third block, which takes account of light quarks at string endpoints, is a tachyon field. It couples to the worldsheet boundary by $S_{\text{q}}=\int d\tau e\,\text{T}$, where $\tau$ is a coordinate on the boundary and $e$ is a boundary metric. In what follows, we consider only the case of $\text{T}(x,r)=\text{T}_0$ and world-sheets whose boundaries are lines in the $t$ direction. In that case, the action written in the static gauge is    

\begin{equation}\label{Sq}
S_{\text q}=\text{T}_0R\int dt \frac{\ep^{\frac{\s}{2}r^2}}{r}
\,,
\end{equation}
and it is nothing else than the action of a point particle of mass ${\text T}_0$ at rest.

\section{The $QQq$-Quark Potential via Gauge/String Duality}
\renewcommand{\theequation}{3.\arabic{equation}}
\setcounter{equation}{0}

Now we will begin our discussion of the ground state of the $QQq$ system. In doing so, we follow the hadro-quarkonium picture and hence think of a light quark as a cloud.\footnote{Clearly, it does not make sense to speak about a light quark position, one can only speak about its average position or, equivalently, about the center of its cloud. We keep that in mind every time we speak about light quarks.} Heavy quarks are point-like objects inside the cloud. It seems natural to suggest that a connected string configuration for the ground state is dictated by symmetry. If so, there are two possible cases: 1. The light quark is in the middle between the heavy quarks. 2. The light quark sits on top of one of the heavy quarks. The latter corresponds to $[Qq]$ diquark formation. 

We start with two connected string configurations that are most symmetric, and then consider a disconnected configuration along with the phenomenon of string breaking.   

\subsection{Connected string configuration I}

\begin{figure}[htbp]
\centering
\includegraphics[width=6.5cm]{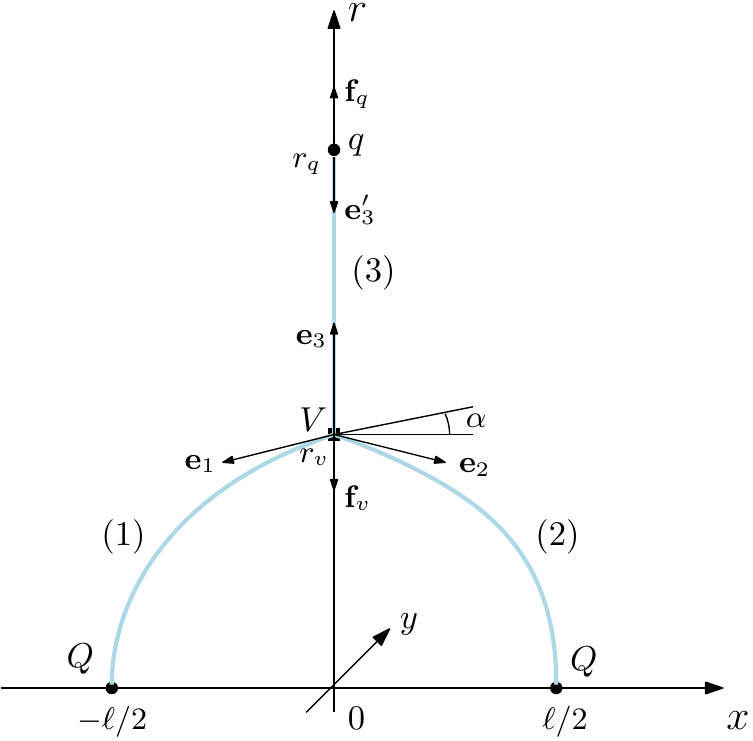}
\caption{{\small A static string configuration at small heavy quark separations. The heavy quarks $Q$ are placed on the $x$-axis, while the light quark $q$ and baryon vertex $V$ on the $r$-axis, respectively at $r=\rq$ and $r=\rv$. The forces exerted on the vertex and light quark are depicted by the arrows.}}
\label{QQqs}
\end{figure}

In the first case, an intuitive way to see the right string configuration is to place the $QQq$ system (like in 4d, three quarks connected by strings) with the light quark exactly in between the heavy quarks on the boundary of five-dimensional space. A gravitational force pulls the light quark and strings into the interior, whereas the heavy (static) quarks remain at rest. As a result, the needed configuration looks like that of Figure \ref{QQqs}. 
Here we consider the case when the static force $\mathbf{f}_v$ pulls the baryon vertex towards the boundary.\footnote{The formal reason for this is that $\tau_v$ turns out to be negative if fitted by the lattice data \cite{a-3q} or by the Lipkin rule. For possible explanations, see Section V.} The configuration is symmetric under the reflection with respect to the $yr$-plane. 
\subsubsection{Small $\ell$}

At small separations between the heavy quarks, a similar configuration appears in the stringy description of triply heavy baryons \cite{a-3q}, as shown in Figure 13 of that paper. The novelty here is that one of the strings terminates on the light quark in the bulk rather than on the third heavy quark on the boundary. 

In the present case, the total action is the sum of the Nambu-Goto actions plus the actions for the vertex and background scalar

\begin{equation}\label{action-s}
S=\sum_{i=1}^3 S_{\text{\tiny NG}}^{(i)}+S_{\text{vert}}+S_{\text q}
\,.
\end{equation}
If one works in the static gauge $\xi^1=t$ and $\xi^2=r$, then the boundary conditions for $x(r)$ become\footnote{We drop the subscript $(i)$ when it does not cause confusion.}

\begin{equation}\label{boundaryc}
x^{(1)}(0)=-\oh\ell\,,\qquad
x^{(2)}(0)=\oh\ell\,,\qquad
x^{(i)}(\rv)=x^{(3)}(\rq)=0\,.
\end{equation}
The action can now be written in more detail as 

\begin{equation}\label{Sconf-s}
S=\g T
\biggl(2\int_{0}^{\rv} \frac{dr}{r^2}\,\ep^{\s r^2}\sqrt{1+(\partial_r x)^2}\,\,+\int_{\rv}^{\rq} \frac{dr}{r^2}\,\ep^{\s r^2}\sqrt{1+(\partial_r x)^2}\,\,
+3\k\,\frac{\ep^{-2\s\rv^2}}{\rv}
+\n\frac{\ep^{\frac{1}{2}\s\rq^2}}{\rq}
\,\biggr)
\,,
\end{equation}
where $\g=\frac{R}{2\pi\alpha'}$, $\k=\frac{\tau_v}{3\g}$, $\n=\frac{\text{T}_0R}{\g}$, $\partial_rx=\frac{\partial x}{\partial r}$, and $T=\int_0^T dt$. To find a stable configuration, one must extremize $S$ with respect to the functions $x(r)$ describing the string profiles and, in addition, with respect to $\rv$ and $\rq$ describing the locations of the baryon vertex and light quark. 

First, we consider the first term in \eqref{Sconf-s} which corresponds to string (1). The equation of motion for $x(r)$ has the form which allows us to immediately integrate it. So, we get 

\begin{equation}\label{calI}
{\cal I}=\frac{w(r)\partial_r x}{\sqrt{1+(\partial_r x)^2}}
\,,\qquad
w(r)=\frac{\ep^{\s r^2}}{r^2}\,.
\end{equation}
Following \cite{a-3q}, we fix ${\cal I}$ by the condition ${\cal I}=\cos\alpha\,w(\rv)$ at $r=\rv$ so that $\partial_r x\vert_{r=\rv}=\cot\alpha$ with $\alpha>0$, as depicted in Figure \ref{QQqs}. Now, by virtue of \eqref{calI}, the integral over $[-\frac{\ell}{2}, 0]$ of $dx$ is equal to

\begin{equation}\label{ls}
\ell=2\cos\alpha\sqrt{\frac{v}{\s}}\int^1_0 du\, u^2\, \ep^{v(1-u^2)}
\Bigl[1-\cos^2{}\hspace{-1mm}\alpha \,u^4\ep^{2v(1-u^2)}\Bigr]^{-\frac{1}{2}}
\,, 
\end{equation}
where $v=\s\rv^2$. 

To compute the string energy, $E=S/T$, we first use the formula \eqref{calI} to express $\partial_rx$ explicitly in terms of $r$ and then, with its help, find that  

\begin{equation}\label{E1Rs}
E_R=\g\sqrt{\frac{\s}{v}}\int^1_{\sqrt{\frac{\s}{v}}\epsilon} \frac{du}{u^2} \ep^{vu^2}
\Bigl[1-\cos^2{}\hspace{-1mm}\alpha \,u^4\ep^{2v(1-u^2)}\Bigr]^{-\frac{1}{2}}
\,.
\end{equation}
Since the integral is divergent at $r=0$ due to the factor $r^{-2}$, in the process we have regularized it by imposing a cutoff $\epsilon$. The regularized expression behaves for $\epsilon\rightarrow 0$ as 

\begin{equation}\label{E1Rs1}
E_R=\frac{\g}{\epsilon}+E+O(\epsilon)\,.
\end{equation}
Subtracting the $\tfrac{1}{\epsilon}$ term and letting $\epsilon=0$, we get a finite result

\begin{equation}\label{E1s}
E=\g\sqrt{\frac{\s}{v}}\int^1_{0} \frac{du}{u^2} 
\biggl(\ep^{vu^2}
\Bigl[1-\cos^2{}\hspace{-1mm}\alpha \,u^4\ep^{2v(1-u^2)}\Bigr]^{-\oh}-1-u^2\biggr)\,\,+c
\,.
\end{equation}
Here $c$ is a normalization constant.  

Since the analysis of string (2) is similar and gives the same result, we jump to string (3) whose action is given by the second term in \eqref{Sconf-s}. It is obvious that the equation of motion derived from this action has a special solution $x=const$ that represents a straight string stretched between the vertex and light quark. The energy evaluated on this solution can be written in the form \cite{a-strb2}

\begin{equation}\label{E3}
	E=\g\int_{\rv}^{\rq}\frac{dr}{r^2}\ep^{\s r^2}
	=
	\g\sqrt{\s}\bigl({\cal Q}(q)-{\cal Q}(v)\bigr)
	\,,
\end{equation}
where $q=\s\rq^2$, ${\cal Q}(x)=\sqrt{\pi}\text{erfi}(\sqrt{x})-\frac{\ep^x}{\sqrt{x}}$, and $\text{erfi}(x)$ is the imaginary error function. The function ${\cal Q}$ is introduced for future use.

Once all the string energies are known, one can easily find the energy of the configuration. From \eqref{Sconf-s}, it follows that 

\begin{equation}\label{EQQqs}
\EI=\g\sqrt{\s}
\biggl(
\frac{2}{\sqrt{v}}\int^1_{0} \frac{du}{u^2} 
\biggl(\ep^{vu^2}
\Bigl[1-\cos^2{}\hspace{-1mm}\alpha \,u^4\ep^{2v(1-u^2)}\Bigr]^{-\oh}-1-u^2\biggr)
\,
+
{\cal Q}(q)-{\cal Q}(v)
+
\n\frac{\ep^{\oh q}}{\sqrt{q}}+3\k\frac{\ep^{-2v}}{\sqrt{v}}\,
\biggr)
+2c\,.
\end{equation}

To complete the analysis, it remains to extremize the action with respect to the positions of the light quark and baryon vertex. This will provide us with gluing recipes describing how to attach the quark to the string endpoint and glue the string endpoints together at the baryon vertex. In conventional language, this means that the net forces exerted on the light quark and vertex must vanish identically. 

We begin with the first case. In that case, the force balance equation is given by 

\begin{equation}\label{fbeq}
	\mathbf{f}_q+\mathbf{e}'_1=0
	\,,
\end{equation}
where $\mathbf{f}_q$ and $\mathbf{e}'_1$ are the forces exerted on the light quark and depicted in Figure \ref{QQqs}. On symmetry grounds, the only non-vanishing components are those in the $r$-direction. By varying the action with respect to $\rq$, we find that $\mathbf{f}_q=(0,-\g\n\,\partial_{r_q}\frac{\ep^{\oh\s\rq^2}}{\rq})$ and $\mathbf{e}_1'=\g w(\rq)(0,-1)$. So, the force balance equation reduces to \footnote{In this form, the parameter $\n$ is expressed in terms of the parameters of \cite{a-strb1} as $\n=\frac{\m}{\g}$.}
 
\begin{equation}\label{fb-q}
\ep^{\frac{q}{2}}+\n(q-1)=0
\,.
\end{equation}
It determines the light quark position. At this point, it is worth mentioning one simple but useful fact about the string tension $\mathbf{e}$: its absolute value is determined by the warp factor so that $\vert \mathbf{e}\vert=\g w(r)$. The above expression for $\mathbf{e}_1'$ is one example of that.

The second case can be analyzed in a similar way. The force balance equation is now  

\begin{equation}\label{fbev}
	\mathbf{e}_1+\mathbf{e}_2+\mathbf{e}_3+\mathbf{f}_v=0
	\,,
\end{equation}
with $\mathbf{f}_v$ and $\mathbf{e}_i$ the forces acting on the vertex, as  shown in Figure \ref{QQqs}. As before, their components in the $r$-direction can be found by varying the action, but this time with respect to $\rv$.\footnote{In doing so, one has to keep in mind the boundary conditions \eqref{boundaryc}.} The remaining components of $\mathbf{e}_1$ and $\mathbf{e}_2$ are determined by the fact that $\vert\mathbf{e}_i\vert=\g w(\rv)$. As a result, we have $\mathbf{f}_v=(0,-3\g\k\,\partial_{\rv}\frac{\ep^{-2\s\rv^2}}{\rv})$, $\mathbf{e}_1=\g w(\rv)(-\cos\alpha,-\sin\alpha)$, $\mathbf{e}_2=\g w(\rv)(\cos\alpha,-\sin\alpha)$, and $\mathbf{e}_3=\g w(\rv)(0,1)$. Thus the force balance equation has only one non-trivial component which can be written as 

\begin{equation}\label{alpha1}
2\sin\alpha-1-3\k(1+4v)\ep^{-3v}=0
\,.	
\end{equation}
This equation determines the angle $\alpha$ at which the strings touch the vertex. 

Thus the energy of the configuration is given in parametric form by $\EI=\EI(v)$ and $\ell=\ell(v)$. The parameter takes values on the interval $[0,q]$.

\subsubsection{Intermediate $\ell$}

\begin{figure}[htbp]
\centering
\includegraphics[width=6.5cm]{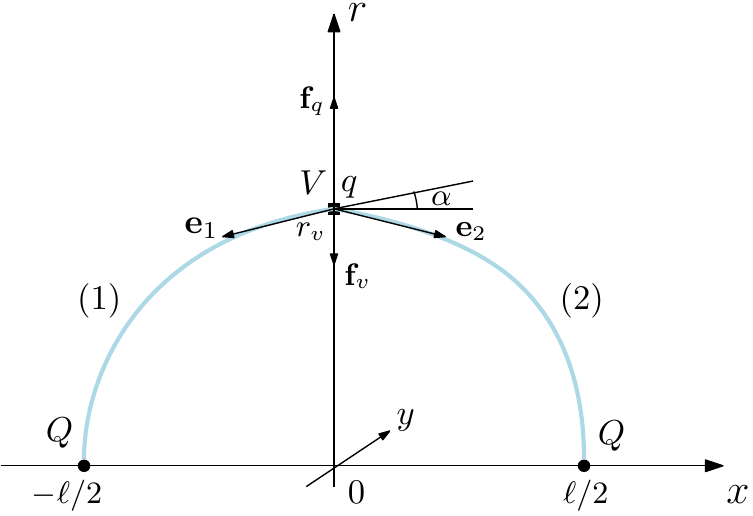}
\caption{{\small A static string configuration at intermediate heavy quark separations. The light quark and vertex reside at the same point on the $r$-axis. The arrows indicate the forces applied at this point.}}
\label{QQqi}
\end{figure}

A simple numerical analysis shows that $\ell(v)$ is an increasing function. Therefore, it reaches its maximum value at $v=q$, which is finite. What happens with the configuration is as follows. As the separation between the heavy quarks is increased, the baryon vertex goes up until it reaches the light quark whose position is independent of the separation. Thus, the string configuration becomes that shown in Figure \ref{QQqi}. One can think 
of it as two strings meeting at a point-like defect made of the light
 quark and vertex. The defect results in a cusp formation in the 
 $r$-direction.\footnote{It is noteworthy that a similar configuration was 
 considered in the context of the hybrid potentials, where a cusp 
 was interpreted as a kind of string excitation \cite{a-hyb}.}
 
It is straightforward to extend the above analysis to the present case. All modifications are due to the absence of string (3). So, the total action now reads

\begin{equation}\label{action-in}
S=\sum_{i=1}^2 S_{\text{\tiny NG}}^{(i)}
+S_{\text{vert}}+S_{\text q}
\,.
\end{equation}

We choose the same static gauge as before. The boundary conditions then take the form 

\begin{equation}\label{bondc-in}
x^{(1)}(0)=-\oh\ell\,,\qquad
x^{(2)}(0)=\oh\ell\,,\qquad
x^{(i)}(\rv)=0\,.
\end{equation}
Clearly, for strings (1) and (2), the expressions \eqref{ls} and \eqref{E1s} hold. So, the energy of the configuration is given by the expression \eqref{EQQqs} but without the contribution from string (3). Explicitly, 

\begin{equation}\label{EQQqm} 
\EI=\g\sqrt{\frac{\s}{v}}
\biggl(
2\int^1_{0} \frac{du}{u^2} 
\biggl(\ep^{vu^2}
\Bigl[1-\cos^2{}\hspace{-1mm}\alpha \,u^4\ep^{2v(1-u^2)}\Bigr]^{-\oh}-1-u^2\biggr)
\,
+
\n\ep^{\oh v}+
3\k\ep^{-2v}\,
\biggr)
+2c\,.
\end{equation}

The force balance equation at the point $r=\rv$ is 

\begin{equation}\label{fbvq}
	\mathbf{e}_1+\mathbf{e}_2+\mathbf{f}_v+\mathbf{f}_q=0
	\,,
\end{equation}
with the forces shown in Figure \ref{QQqi}. The $r$-components of these vectors are again found by varying the total action with respect to $\rv$. The remaining $x$-components of the $\mathbf{e}$'s are determined by the condition $\vert\mathbf{e}_i\vert=\g w(\rv)$. This gives $\mathbf{f}_q=(0,-\g\n\,\partial_{r_q}\frac{\ep^{\oh\s\rq^2}}{\rq})$, $\mathbf{f}_v=(0,-3\g\k\,\partial_{\rv}\frac{\ep^{-2\s\rv^2}}{\rv})$, $\mathbf{e}_1=\g w(\rv)(-\cos\alpha,-\sin\alpha)$, $\mathbf{e}_2=\g w(\rv)(\cos\alpha,-\sin\alpha)$. The equation therefore simplifies to

\begin{equation}\label{alpha2}
2\sin\alpha-3\k(1+4v)\ep^{-3v}+\n (v-1)\ep^{-\oh v}=0
\,.
\end{equation}
This equation determines the angle $\alpha$. Since our analysis is only valid for non-negative values of $\alpha$, this imposes a restriction on the range of allowed values of $v$. The upper limit $v_0$ corresponds to $\alpha=0$. Therefore it is a solution to

\begin{equation}\label{v0}
3\k(1+4v)+\n (1-v)\ep^{\frac{5}{2}v}=0
\end{equation}
in the interval $[q,1]$. The value $v_0$ has a clear meaning. For this value of $v$, the point $r=\rv$ becomes a smooth point, not a cusp. So the strings joint smoothly. 

In summary, at intermediate separations the energy $\EI$ is given by the parametric equations \eqref{ls} and \eqref{EQQqm}, with $\alpha$ defined by \eqref{alpha2}. The parameter $v$ takes values in the interval $[q,v_0]$. 

\subsubsection{Large $\ell$}

\begin{figure}[htbp]
\centering
\includegraphics[width=6.95cm]{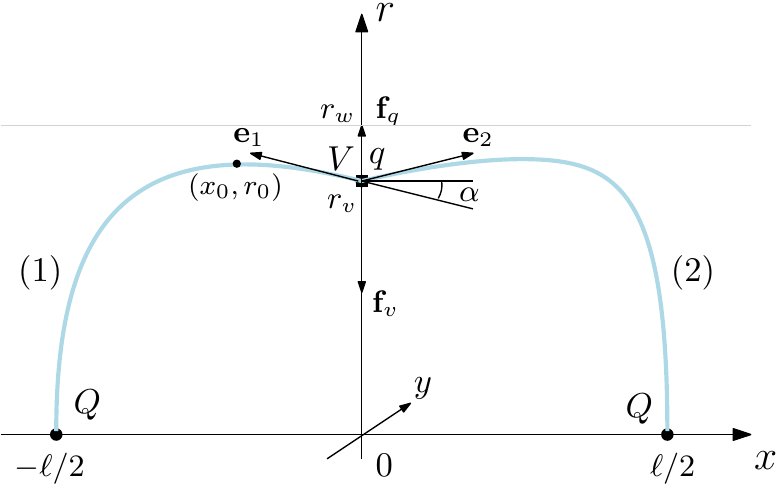}
\caption{{\small A static configuration at large heavy quark separations. For string (1) a turning point is located at $(x_0,r_0)$. The soft wall is at $r_w=1/\sqrt{\s}$. The arrows indicate the forces applied at the point $r=\rv$.}}
\label{QQql}
\end{figure}

From Eq.\eqref{ls}, it follows that $\ell$ remains finite and bounded from above by $\ell(v_0)$.\footnote{The argument assumes that $\ell(v)$ is an increasing function of $v$. This is indeed the case for the model parameter values we use.} So, the question arises what is going to happen for larger values of $\ell$? The short answer to this question is that $\alpha$ changes the sign from positive to negative so that the configuration profile becomes convex near $x=0$, as the one depicted in Figure \ref{QQql}. This situation continues until the strings finally reach the soft-wall that corresponds to infinite separation between the heavy quarks.  

The configuration is again governed by the total action \eqref{action-in}. But now it is more convenient to choose another static gauge such that $\xi^1=t$ and $\xi^2=x$. In this gauge, the boundary conditions are written as

\begin{equation}\label{bondc-l}
r^{(1)}(-\ell/2)=r^{(2)}(\ell/2)=0\,,\qquad
r^{(i)}(0)=\rv\,.
\end{equation}
The total action is thus 

\begin{equation}\label{conf-s}
S=\g T\biggl(
\int_{-\ell/2}^{0} dx\,w(r)\sqrt{1+(\partial_x r)^2}+
\int_{0}^{\ell/2} dx\,w(r)\sqrt{1+(\partial_x r)^2}\,
+3\k\,\frac{\ep^{-2\s\rv^2}}{\rv}
+\n\frac{\ep^{\frac{1}{2}\s\rv^2}}{\rv}\,
\biggr)
\,.
\end{equation}

To start with, we consider string (1) whose action is given by the first term in \eqref{conf-s}.\footnote{More detail can be found in Appendix A of \cite{a-3q}.} Since the integrand does not explicitly depend on $x$, the equation of motion has the first integral 

\begin{equation}\label{calI2}
{\cal I}=\frac{w(r)}{\sqrt{1+(\partial_x r)^2}}
\,.
\end{equation}
There is nothing unusual about this integral, it is the same integral as in \eqref{calI}. ${\cal I}$ is equal to $w(r_0)$ at the turning point and to 
$w(\rv)\cos\alpha$, with $\alpha<0$, at the string endpoint. From this, it follows that 

\begin{equation}\label{lambda-v}
\frac{\ep^{\lambda}}{\lambda}=\frac{\ep^{v}}{v}\cos\alpha	
\,,
\end{equation}
where $\lambda=\s x_0^2$. 

If we split the interval into two subintervals at the turning point, and then integrate the differential equation \eqref{calI2} over both subintervals, we get 

\begin{equation}\label{l1}
x_0+\oh\ell=\sqrt{\frac{\lambda}{\s}}
\int^1_0 du\, u^2\, \ep^{\lambda(1-u^2)}
\Bigl[1-u^4\,\ep^{2\lambda(1-u^2)}\Bigr]^{-\frac{1}{2}}
\,
\end{equation}
on the first subinterval, and 

\begin{equation}\label{l2}
x_0=-\sqrt{\frac{\lambda}{\s}}
\int^1_{\sqrt{\frac{v}{\lambda}}} du\, u^2\, \ep^{\lambda(1-u^2)}
\Bigl[1-u^4\,\ep^{2\lambda(1-u^2)}\Bigr]^{-\frac{1}{2}}
\,
\end{equation}
on the second. Combining them gives 

\begin{equation}\label{l3}
\ell=2\sqrt{\frac{\lambda}{\s}}
\biggl(
\int^1_0 du\, u^2\, \ep^{\lambda(1-u^2)}
\Bigl[1-u^4\,\ep^{2\lambda(1-u^2)}\Bigr]^{-\frac{1}{2}}
+
\int^1_{\sqrt{\frac{v}{\lambda}}} du\, u^2\, \ep^{\lambda(1-u^2)}
\Bigl[1-u^4\,\ep^{2\lambda(1-u^2)}\Bigr]^{-\frac{1}{2}}
\biggr)
\,.
\end{equation}
The integrals are well-defined as long as $\lambda<1$, or in other words the string does not touch the soft-wall.  

The string energy can be computed by first splitting the interval at the turning point, as in the previous case, and then using the first integral \eqref{calI2} to express the integrand in terms of $r$ only. Since the resulting integral diverges at $r=0$,  we regularize it by imposing a cutoff $r\geq \epsilon$. In this way we arrive at 

\begin{equation}\label{E1-1}
E_{R}=\sqrt{\frac{\s}{\lambda}}
\biggl(
\int^1_{\sqrt{\frac{\s}{\lambda}}\epsilon} \frac{du}{u^2}\, \ep^{\lambda u^2}
\Bigl[1-u^4\,\ep^{2\lambda(1-u^2)}\Bigr]^{-\frac{1}{2}}
+
\int^1_{\sqrt{\frac{v}{\lambda}}}\frac{du}{u^2}\, \ep^{\lambda u^2}
\Bigl[1-u^4\,\ep^{2\lambda(1-u^2)}\Bigr]^{-\frac{1}{2}}
\biggr)
\,.
\end{equation}
After subtracting the $\frac{1}{\epsilon}$ term and letting $\epsilon=0$, we get a finite result for the energy of string (1)

\begin{equation}\label{E1-2}
E=\sqrt{\frac{\s}{\lambda}}
\biggl(
\int^1_{0} \frac{du}{u^2}\, 
\Bigl(\ep^{\lambda u^2}
\Bigl[1-u^4\,\ep^{2\lambda(1-u^2)}\Bigr]^{-\frac{1}{2}}-1-u^2\Bigr)
+
\int^1_{\sqrt{\frac{v}{\lambda}}}\frac{du}{u^2}\, \ep^{\lambda u^2}
\Bigl[1-u^4\,\ep^{2\lambda(1-u^2)}\Bigr]^{-\frac{1}{2}}
\biggr)\,+c
\,.
\end{equation}

The analysis of string (2) proceeds in a similar way and gives the same expressions for $\ell$ and $E$, as expected from the symmetry arguments. Once the energies of the strings are known, the energy of the configuration can be written as

\begin{equation}\label{EQQql}
\begin{split}
\EI=&2\g\sqrt{\frac{\s}{\lambda}}
\biggl(
\int^1_{0} \frac{du}{u^2} 
\Bigl(\ep^{\lambda u^2}
\Bigl[1-u^4\ep^{2\lambda(1-u^2)}\Bigr]^{-\frac{1}{2}}-1-u^2\Bigr)
+
\int^1_{\sqrt{\frac{v}{\lambda}}} \frac{du}{u^2} 
\ep^{\lambda u^2}
\Bigl[1-u^4\ep^{2\lambda(1-u^2)}\Bigr]^{-\frac{1}{2}}\biggr)\\
+&\g\sqrt{\frac{\s}{v}}
\Bigl(\n\ep^{\oh v}+3\k\ep^{-2v}\Bigr)
+2c\,.
\end{split}
\end{equation}
 
The force balance equation at the point $r=\rv$ is given by Eq.\eqref{fbvq} with the same expressions for the components of each force, but now with a negative value of $\alpha$. So it simplifies to Eq.\eqref{alpha2} again. The parameter $v$ runs from $v_0$ to $v_1$, where $v_1$ is a solution to

\begin{equation}\label{v1}
2\sqrt{1-v^2\ep^{2(1-v)}}+3\k(1+4v)\ep^{-3v}+\n (1-v)\ep^{-\oh v}=0
\,.
\end{equation}
The meaning of this solution is as follows. It corresponds to $\lambda=1$. In that case the strings approach the soft wall and the separation distance between the heavy quarks becomes infinite. 

Thus, at large separations the energy of the configuration is given in parametric form by $\EI=\EI(v)$ and $\ell=\ell(v)$. The parameter $v$ takes values on the interval $[v_0,v_1]$. Note that $\lambda(v)$ can be expressed explicitly in terms of the ProductLog function \cite{wolf}

\begin{equation}
\lambda(v)=-\text{ProductLog}\Bigl[-v\ep^{-v}
\Bigl(1-\frac{1}{4}\Bigl(3\k(1+4v)\ep^{-3v}
+
\n(1-v)\ep^{-\oh v}\Bigr)^2
\,\Bigr)^{-\frac{1}{2}}
\,\Bigl]
\,,
\end{equation}	
as seen from \eqref{alpha2} and \eqref{lambda-v}. 

As such, the main conclusion of our analysis is that $\EI$ is a piecewise function of $\ell$. The point is that the shape of this string configuration depends on the separation between the heavy quarks.

\subsubsection{More on the limiting cases}

Having derived the parametric formulas for $\ell$ and $\EI$, it is instructive to examine the behavior of $\EI$ at small and large separations in order to see some of the main features of our model. 

We begin with the case of small $\ell$. Because $\ell$ is an increasing function of $v$ on the whole interval $[0,v_1]$, small $v$'s correspond to small $\ell$'s. From \eqref{ls}, the behavior of $\ell$ near $v=0$ is 

\begin{equation}\label{l-small}
\ell=\sqrt{\frac{v}{\s}}\Bigl(l_0+l_1v+O(v^2)\Bigl)
\,,
\end{equation}
where $l_0=\frac{1}{2}\xi^{-\frac{1}{2}}B\bigl(\xi^2;\tfrac{3}{4},\tfrac{1}{2}\bigr)$ and $l_1=\frac{1}{2}\xi^{-\frac{3}{2}}
\bigl[ \bigl(2\xi+\frac{3}{4}\frac{\k-1}{\xi}\bigr)B\bigl(\xi^2;\tfrac{3}{4},-\tfrac{1}{2}\bigr)-B\bigl(\xi^2;\tfrac{5}{4},-\tfrac{1}{2}\bigr)\bigr]$. Here $\xi=\frac{\sqrt{3}}{2}(1-2\k-3\k^2)^{\frac{1}{2}}$ and $B(z;a,b)$ is the incomplete beta function. The corresponding expansion for the energy is obtained from \eqref{EQQqs}. Expanding up to the quadratic terms in $v$, we arrive at

\begin{equation}
\EI=\g\sqrt{\frac{\s}{v}}\Bigl(E_0+E_1 v+O(v^2)\Bigr) +
E_{\qQb}+c
\,,
\end{equation}
with $E_0=1+3\k+\frac{1}{2}\xi^{\frac{1}{2}}B\bigl(\xi^2;-\tfrac{1}{4},\tfrac{1}{2}\bigr)$ and $E_1=\xi\,l_1-1-6\k+\frac{1}{2}\xi^{-\frac{1}{2}}B\bigl(\xi^2;\tfrac{1}{4},\tfrac{1}{2}\bigr)$. The constant term $E_{\qQb}$ coincides with $E_{\Qqb}$ of \cite{a-strb1} which was interpreted there as a heavy meson mass in the static limit. Explicitly,

\begin{equation}\label{EQqb}
E_{\Qqb}=\g\sqrt{\s}\Bigl({\cal Q}(q)+\n\frac{\ep^{\oh q}}{\sqrt{q}}\,\Bigr)+c
\,.
\end{equation}
	
These two equations reduce to a single equation in which the parameter $v$ is absent 

\begin{equation}\label{E-small}
\EI=-\frac{\alpha_{\QQq}}{\ell}+E_{\qQb}+c+\boldsymbol{\sigma}_{\QQq}\ell+O(\ell^2)
\,,
\end{equation}
where $\alpha_{\QQq}=-l_0E_0\g$ and $\boldsymbol{\sigma}_{\QQq}=\frac{1}{l_0}\Bigl(E_1+\frac{l_1}{l_0}E_0\Bigr)\g\s$. Importantly, these coefficients are the same as those for the heavy quark-quark potential \cite{a-3q} with the quarks in a color antitriplet state, namely $\alpha_{\QQq}=\alpha_{\QQ}$ and $\boldsymbol{\sigma}_{\QQq}=\boldsymbol{\sigma}_{\QQ}$. If so, then Eq.\eqref{E-small} tells us that  

\begin{equation}\label{factor}
\EI(\ell)=E_{\qQb}+E_{\QQ}(\ell)\,,
\qquad\text{with}
\qquad 
E_{\QQ}=-\frac{\alpha_{\QQ}}{\ell}+c+\boldsymbol{\sigma}_{\QQ}\ell
\,.
\end{equation}	
Thus the model we are considering does have the desired property of factorization, expected from heavy quark-diquark symmetry \cite{wise}. We should stress that what is important in \eqref{factor} is that it is valid up to linear terms in $\ell$.

In a similar spirit, we can explore the large $\ell$ behavior of $\EI$. In the limit $v\rightarrow v_1$, or equivalently $\lambda\rightarrow 1$, the strings become infinitely long, as follows from Eq.\eqref{l3}. Moreover, both integrals in \eqref{EQQql} approach infinity, while the other terms remain finite. First we compute the leading term. The strategy of the computation is very similar to that of \cite{az1}. The behavior of $\ell$ and $\EI$ near $\lambda=1$ is given by 

\begin{equation}\label{sigular}
\ell(\lambda)=-\frac{2}{\sqrt{\s}}\ln(1-\lambda)+O(1)
\,,\qquad
\EI(\lambda)=-2\g\ep\sqrt{\s}\ln(1-\lambda)+O(1)
\,.
\end{equation}
From this, it follows that  

\begin{equation}\label{linear-l}
\EI=\sigma\ell +O(1)\,,
\qquad \text{with}\qquad
\sigma=\g\ep\s
\,.	
\end{equation}
Here $\sigma$ is the physical string tension. Thus, the important property of the model is that the string tension is universal. It is the same in all the cases: the quark-antiquark \cite{az1}, hybrid \cite{a-hyb}, three-quark potentials \cite{a-3q0}, and also in the present example of the $QQq$ potential.  

To find the next-to-leading term in the large $\ell$ expansion, consider 

\begin{equation}\label{diff}
\begin{split}
\EI-\sigma\ell=&2\g\sqrt{\frac{\s}{\lambda}}
\biggl(
\int_0^1\frac{du}{u^2}
\Bigl(\ep^{\lambda u^2}
\Bigl[1-\lambda u^4\ep^{1+\lambda(1-2u^2)}\Bigr]
\Bigl[1-u^4\ep^{2\lambda(1-u^2)}\Bigr]^{-\frac{1}{2}}
-1-u^2\Bigr)\,
\\
+&
\int_{\sqrt{\frac{v}{\lambda}}}^1\frac{du}{u^2}\ep^{\lambda u^2}
\Bigl[1-\lambda u^4\ep^{1+\lambda(1-2u^2)}\Bigr]
\Bigl[1-u^4\ep^{2\lambda (1-u^2)}\Bigr]^{-\frac{1}{2}}
+
\oh\sqrt{\frac{\lambda}{v}}
\Bigl(
\n\ep^{\oh v}+3\k\ep^{-2 v}\Bigr)\,
\biggr)+2c
\,.
\end{split}
\end{equation}
The result of taking $v\rightarrow v_1$ gives 

\begin{equation}\label{diff2}
\EI-\sigma\ell=2\g\sqrt{\s}
\biggl(
\int_0^1\frac{du}{u^2}
\Bigl(\ep^{u^2}\Bigl[1-u^4\ep^{2(1-u^2)}\Bigr]^{\frac{1}{2}}-1-u^2\Bigr)
+
\int_{\sqrt{v_1}}^1\frac{du}{u^2}\ep^{u^2}\Bigl[1-u^4\ep^{2(1-u^2)}\Bigr]^{\frac{1}{2}}
+
\frac{\n\ep^{\oh v_1}+3\k\ep^{-2 v_1}}{2\sqrt{v_1}}
\biggr)+2c
\,.
\end{equation}
 From this, it follows that 

\begin{equation}\label{EQQq-large}
	\EI=\sigma\ell-2\g\sqrt{\s}\II+2c+o(1)
	\,,
	\end{equation}
where 
\begin{equation}\label{IQQq}
	\II=
I_0
-
\int_{\sqrt{v_1}}^1\frac{du}{u^2}\ep^{u^2}\Bigl[1-u^4\ep^{2(1-u^2)}\Bigr]^{\frac{1}{2}}
-
\frac{\n\ep^{\oh v_1}+3\k\ep^{-2 v_1}}{2\sqrt{v_1}}
\,.
\end{equation}
Here $I_0$ is a non-zero constant defined in the Appendix. 


\subsubsection{Some miscellaneous remarks}

We conclude our discussion of configuration I with some miscellaneous remarks.

\begin{figure}[htbp]
\centering
\includegraphics[width=8.2cm]{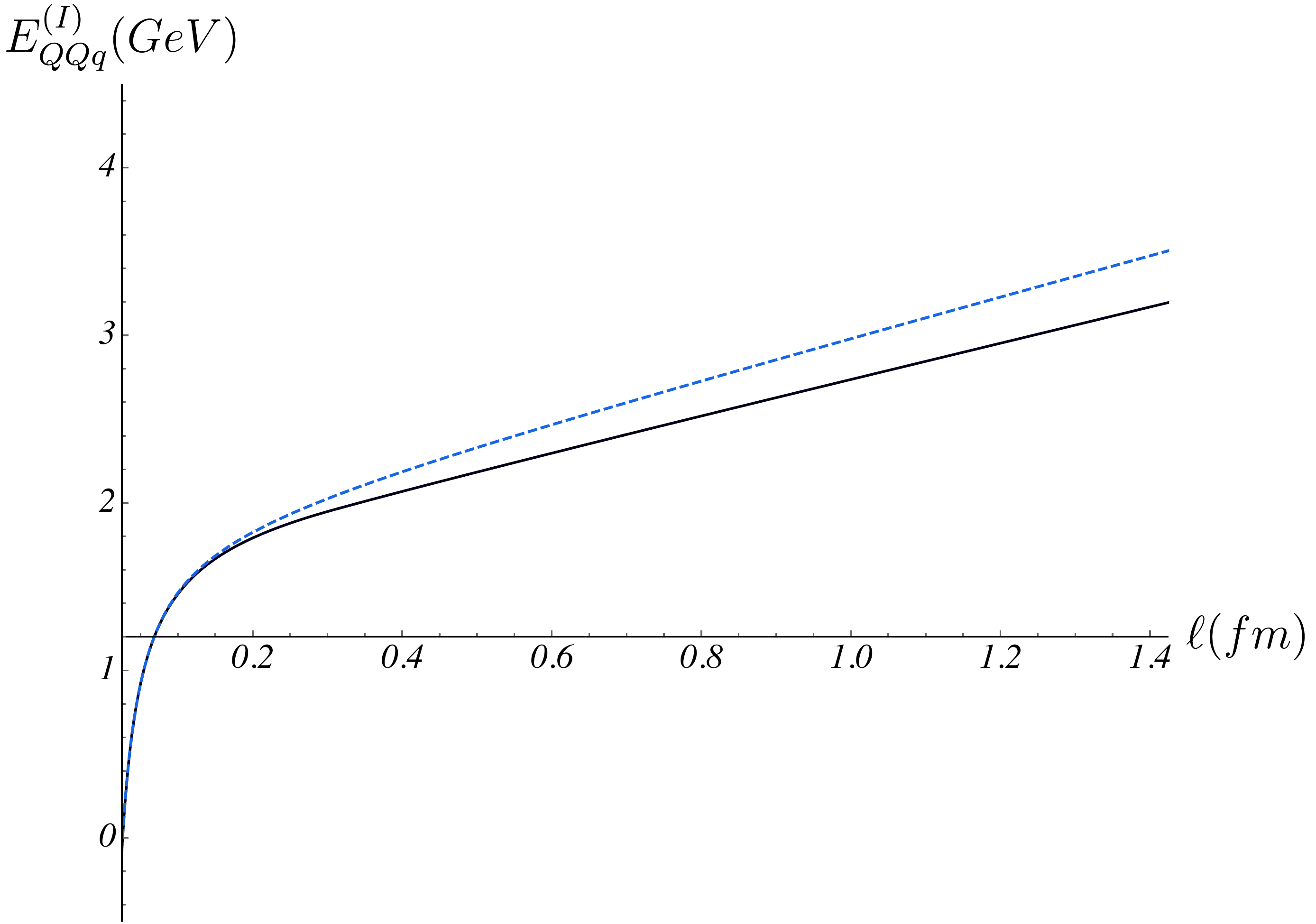}
\hspace{1.25cm}
\includegraphics[width=8.2cm]{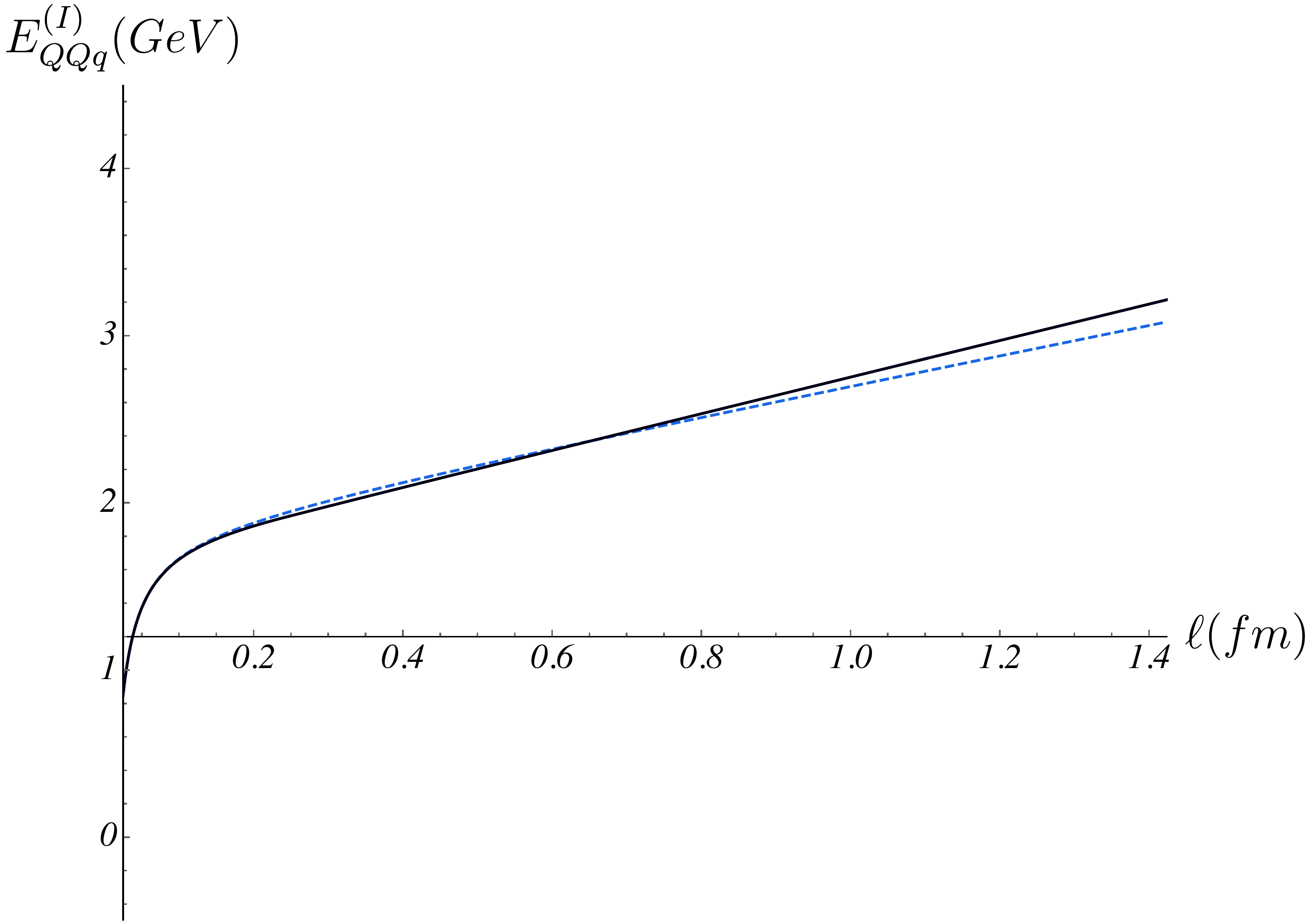}
\caption{{\small $\EI$ as a function of $\ell$ at $\k=-\frac{1}{4}\ep^{\frac{1}{4}}$ (left panel) and $\k=-0.102$ (right panel). The dashed curves represent the small $\ell$ behavior, as obtained from Eq.\eqref{E-small}.}}
\label{E2QqI}
\end{figure}

First of all, it is instructive to plot the piecewise function $\EI(\ell)$ to see how well the pieces align. In doing so, we use the set of parameters denoted by $L$ in \cite{a-strb1}. It follows closely lattice QCD and includes $\s=0.45\,\text{GeV}^2$, $\g=0.176$, and $\n=3.057$ along with $c=0.623\,\text{GeV}$. Examples of $\EI$ for two different values of $\k$ are given in Figure \ref{E2QqI}.\footnote{This requires a caveat. In general, this string model gives reliable results only for separations larger than $0.1\,\text{fm}$. } 
We see that the function is smoothly increasing, as expected. However, we observe that the upper limit of the interval, where $\EI$ is well approximated by the asymptotic formula \eqref{E-small}, strongly depends on the value of $\k$. It is about $0.2\,\text{fm}$ at $k=-\frac{1}{4}\ep^{\frac{1}{4}}$, which is approximately $-0.321$, and then it goes up to $0.8\,\text{fm}$ at $\k=-0.102$.\footnote{Note that $\k=-0.102$ is a numerical solution of equation $\alpha_{\QQ}(\k)=\oh\alpha_{\QQb}$. We will discuss this later in Sec.IV.} This implies that heavy quark-diquark symmetry could be a reasonable approximation up to such long distances. Another observation we would like to make here is that $\EI$ becomes a linear function of $\ell$ already for small interquark separations. As seen from the Figure, this is true for $\ell\gtrsim 0.25\,\text{fm}$ independently on the value of $\k$.

\begin{figure}[htbp]
\centering
\includegraphics[width=6.5cm]{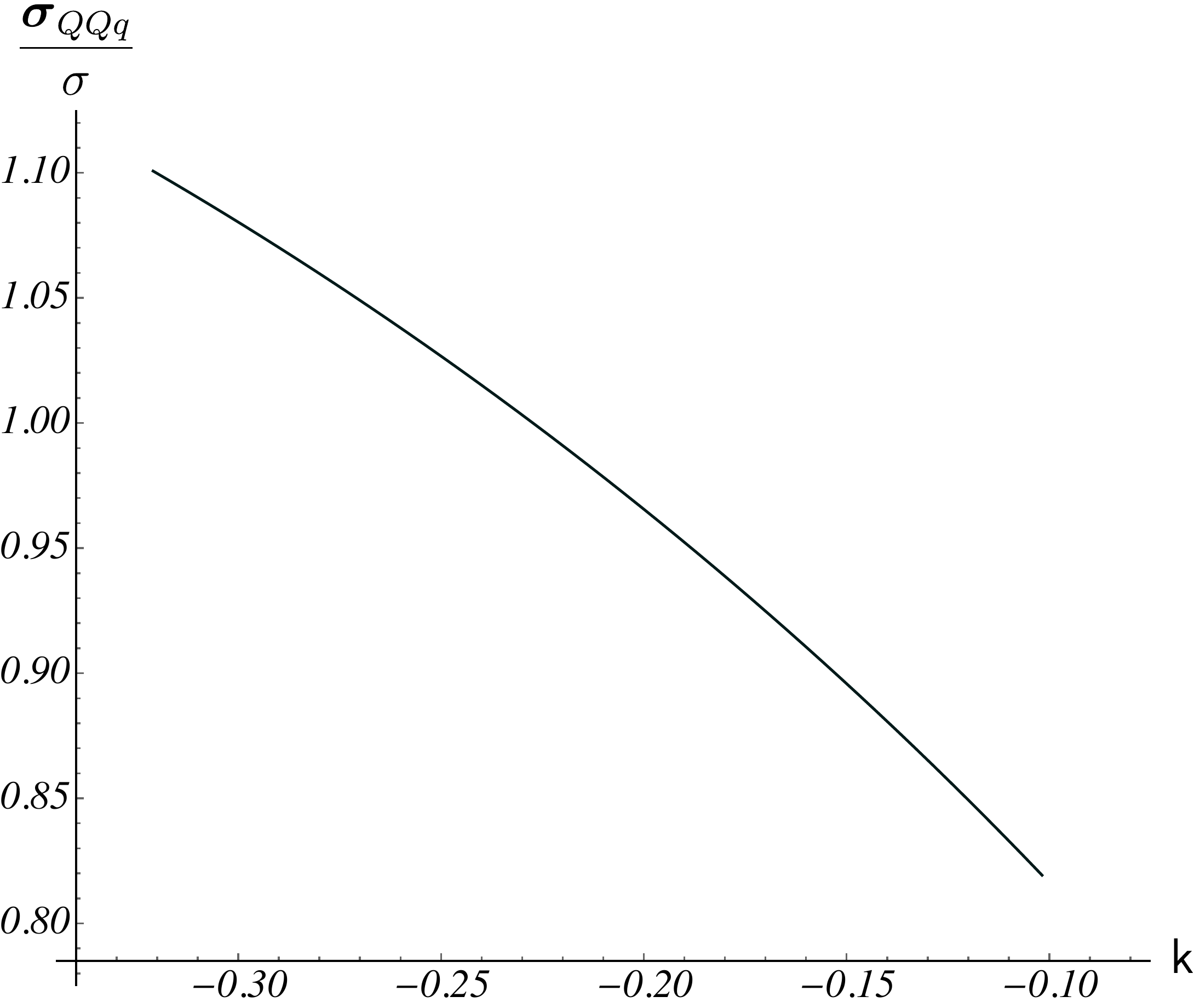}
\hspace{2cm}
\includegraphics[width=6.75cm]{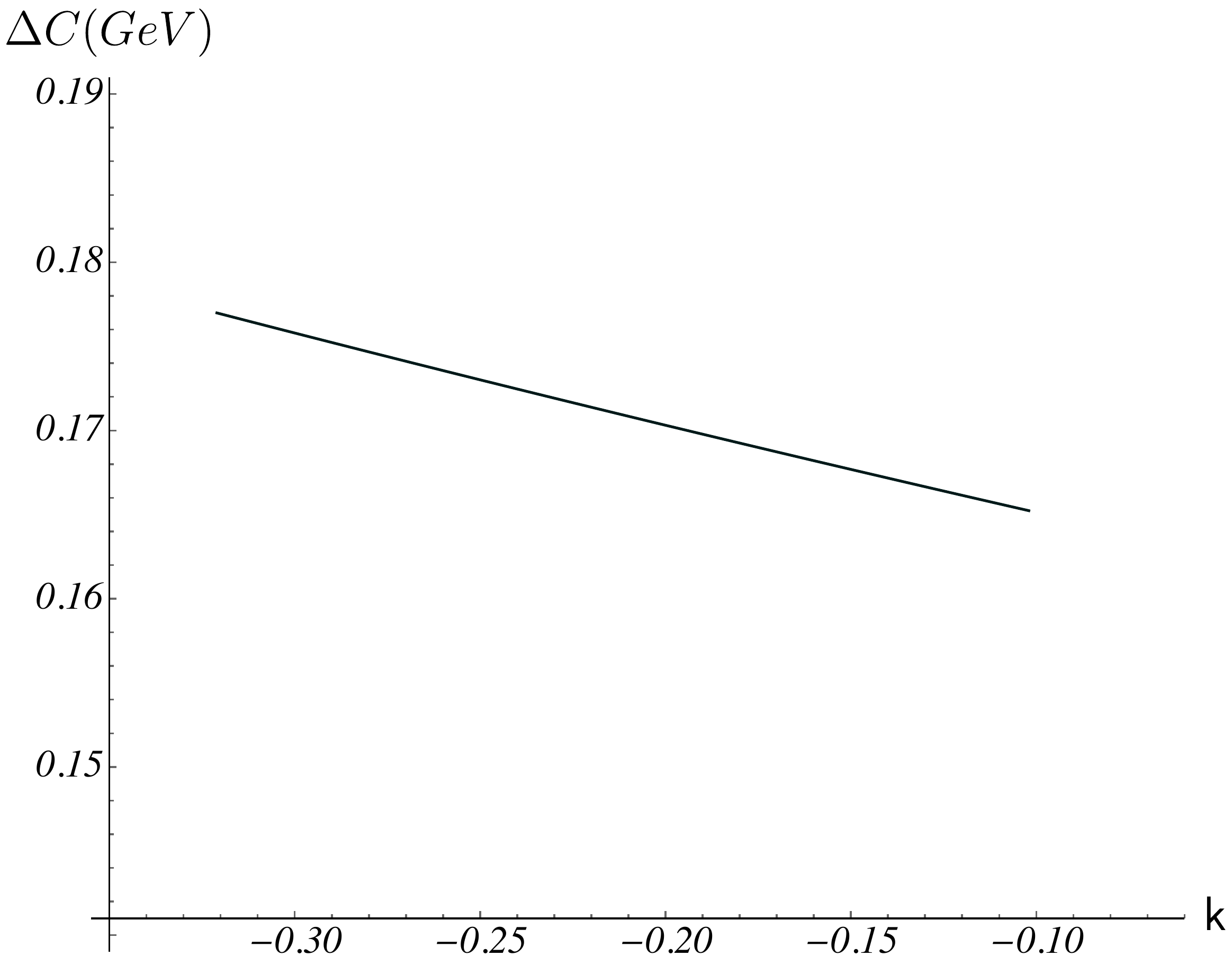}
\caption{{\small Left: Tension $\boldsymbol{\sigma}_{\QQq}$ in units of $\sigma$ versus $\k$. Right: $\Delta C$ versus $\k$ for the parameter set $L$. In both cases, $\k$ runs from $-0.321$ to $-0.102$.}}
\label{sigmak}
\end{figure}

Although the $QQq$-potential was studied in lattice QCD \cite{suganuma,bali}, the lattice data are limited to interquark separations not exceeding $0.8\,\text{fm}$,  with the Coulomb coefficient $\alpha_{\QQq}$ is almost equal to $\oh\alpha_{\QQb}$.   In the model we are considering, the latter implies that the value of $\k$ is close enough to $-0.102$, as shown in the right panel of Figure \ref{E2QqI}. If so, then the data could be well-fitted by Eq.\eqref{E-small}, which is exactly the functional form used in \cite{suganuma,bali}. This makes it difficult to extract the physical string tension, not $\boldsymbol{\sigma}_{\QQq}$. There is more evidence in favor of such a scenario. The effective tension, as claimed in \cite{suganuma}, is about $10$-$20\%$ less than the physical string tension. As seen from the left panel of Figure \ref{sigmak}, $\boldsymbol{\sigma}_{\QQq}$ is in this range if $\k$ near $-0.102$. 
Of course, a natural way to resolve this issue in lattice QCD would be to probe larger separations which are better suited for extracting the physical string tension. 

In the string model we are considering, one observation that makes it different from the majority of phenomenological models is that the constant terms are different in the small and large $\ell$ expansions of $E(\ell)$'s. We can also see this for $E_{\QQq}(\ell)$. In this case, the difference between the constant terms can be read off from Eqs.\eqref{E-small} and \eqref{EQQq-large}. So, we have 

\begin{equation}\label{Delta-C}
\Delta C= \g\sqrt{\s}\Bigl({\cal Q}(q)+\n\frac{\ep^{\oh q}}{\sqrt{q}}+2\II\Bigr)
\,.
\end{equation}
It is finite and scheme-independent. Moreover, it turns out that in all the known examples the constant terms at small separations are larger than those at large separations \cite{az1,a-hyb,a-3q}. As seen from the right panel of Figure \ref{sigmak}, this is also true in the present case. 

\subsection{Connected string configuration II}

Having understood the symmetric string configuration, we can now consider another configuration which is not symmetric but contains a diquark structure $[Qq]$. It somehow mimics the situation when the light quark sits on top of one of the heavy quarks. 

\subsubsection{A construction}
\begin{figure}[htbp]
\centering
\includegraphics[width=6.8cm]{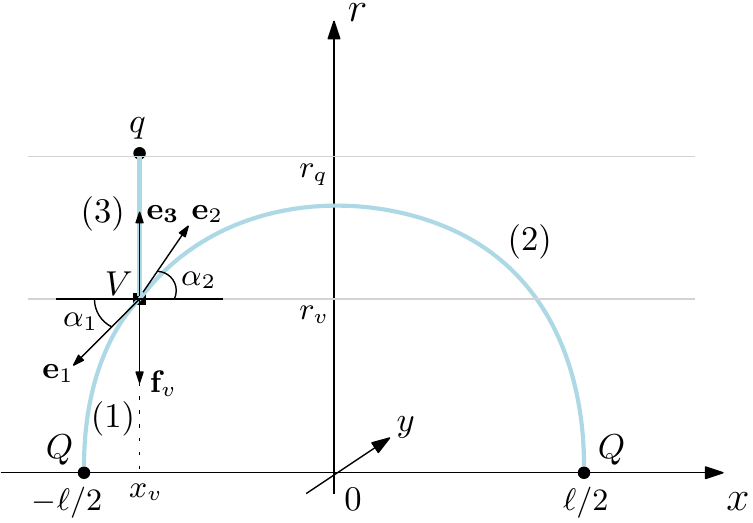}
\caption{{\small A non-symmetric static configuration. The baryon vertex is located on the $xr$-plane at the point $(x_v,r_v)$.}}
\label{QQqd}
\end{figure}

If the reflection symmetry is broken, then the first that comes to mind is to consider a string configuration as that shown in Figure \ref{QQqd}. It is governed by the 
same action that we originally encountered in the case of configuration I for small heavy quark separations. However, in the static gauge $\xi^1=t$ and $\xi^2=r$, the boundary conditions are replaced by

\begin{equation}\label{bcond-d}
x^{(1)}(0)=-\oh\ell\,,\qquad
x^{(2)}(0)=\oh\ell\,,\qquad
x^{(i)}(\rv)=x^{(3)}(\rq)=x_v\,.
\end{equation}
The corresponding action reads 

\begin{equation}\label{ac-d}
\begin{split}
S=\g T
\biggl(
&\int_{0}^{\rv} \frac{dr}{r^2}\,\ep^{\s r^2}\sqrt{1+(\partial_r x)^2}
+\int_{0}^{\rq} \frac{dr}{r^2}\,\ep^{\s r^2}\sqrt{1+(\partial_r x)^2}
+\int_{\rv}^{\rq} \frac{dr}{r^2}\,\ep^{\s r^2}\sqrt{1+(\partial_r x)^2}\\
+&\int_{\rv}^{\rq} \frac{dr}{r^2}\,\ep^{\s r^2}
+3\k\,\frac{\ep^{-2\s\rv^2}}{\rv}
+\n\frac{\ep^{\frac{1}{2}\s\rq^2}}{\rq}
\,\biggr)
\,.
\end{split}
\end{equation}
Here the first integral corresponds to string (1), the next two integrals to string (2), and the last to string (3) with $x=const$. 

We begin with the force balance equations. They are again given by equations of the form \eqref{fbeq} and \eqref{fbev}. Clearly, nothing happens with that equation for the light quark. It reduces to Eq.\eqref{fb-q}. By contrast, some forces exerted on the baryon vertex are altered to shift the balance, as depicted schematically in Figure \ref{QQqd}. These are $\mathbf{e}_1$ and $\mathbf{e}_2$, whose components become $\mathbf{e}_1=\g w(\rv)(-\cos\alpha_1,-\sin\alpha_1)$ and $\mathbf{e}_2=\g w(\rv)(\cos\alpha_2,\sin\alpha_2)$ with $\alpha_i\leq\frac{\pi}{2}$. If so, then the $x$-component of Eq.\eqref{fbev} results in 

\begin{equation}\label{fb-x}
\cos\alpha_1-\cos\alpha_2 =0
\,. 
\end{equation}
This equation has a trivial solution $\alpha_1=\alpha_2$. "Trivial" means that in this case the strings (1) and (2) are smoothly glued together to form a single string. Another consequence of this solution is that the $r$-component is simply 

\begin{equation}\label{fbv2}
1+3\k(1+4v)\ep^{-3v}=0
\,.
\end{equation}
This equation determines the position of the baryon vertex in the bulk. It is a special case of Eq.\eqref{alpha1} in which $\alpha=0$. It is noteworthy that the position does not depend on the separation between the heavy quarks. Thus we have learned that in the case of configuration II the force balance equation \eqref{fbev} splits into two pieces. One piece is $\mathbf{e}_1+\mathbf{e}_2=0$ and the other is $\mathbf{f}_v+\mathbf{e}_3=0$. This allows one to treat strings (1) and (2) as a single string. 

Before we continue with the discussion, let us pause here to make a couple of remarks about equation \eqref{fbv2}. A simple analysis shows that in the interval $(0,1)$ this equation has solutions if $\k$ is restricted to the range $-\frac{\ep^3}{15} <\k\leq -\frac{1}{4}\ep^{\frac{1}{4}}$. In particular, there exists a single solution $v=\frac{1}{12}$ at $\k= -\frac{1}{4}\ep^{\frac{1}{4}}$. In order that the configuration is stable, the condition $v\leq q$ must be satisfied. This may further restrict the range of allowed values of $\k$. 

Having understood the force balance equations, it is straightforward to write down the energy of the configuration. But first let us mention, that the energy of a string stretched between two heavy quark sources placed on the boundary of five-dimensional space can be written parametrically as $E_{\QQb}=E_{\QQb}(\lambda)$ and $\ell=\ell(\lambda)$, with $\lambda$ a parameter (see the Appendix). Now combining the first three terms in \eqref{ac-d} into $E_{\QQb}$ and performing the remaining integral, we find

\begin{equation}\label{E2Qq-d}
	\EII=E_{\QQb}+\g\sqrt{\s}\Bigl({\cal Q}(q)-{\cal Q}(\vv)+\n\frac{\ep^{\oh q}}{\sqrt{q}}+3\k\frac{\ep^{-2\vv}}{\sqrt{\vv}}
	\Bigr)\,,
\end{equation}
where $E_{\QQb}$ is given by \eqref{EQQb} and $\vv$ is a solution of  Eq.\eqref{fbv2}. Importantly, the parameter $\lambda$ runs from $\vv$ to $1$. Thus, such a connected configuration does not exist for $\ell$ smaller than $\ell(\vv)$.

\subsubsection{Behavior of $\EII$}

\begin{figure}[htbp]
\centering
\includegraphics[width=7.85cm]{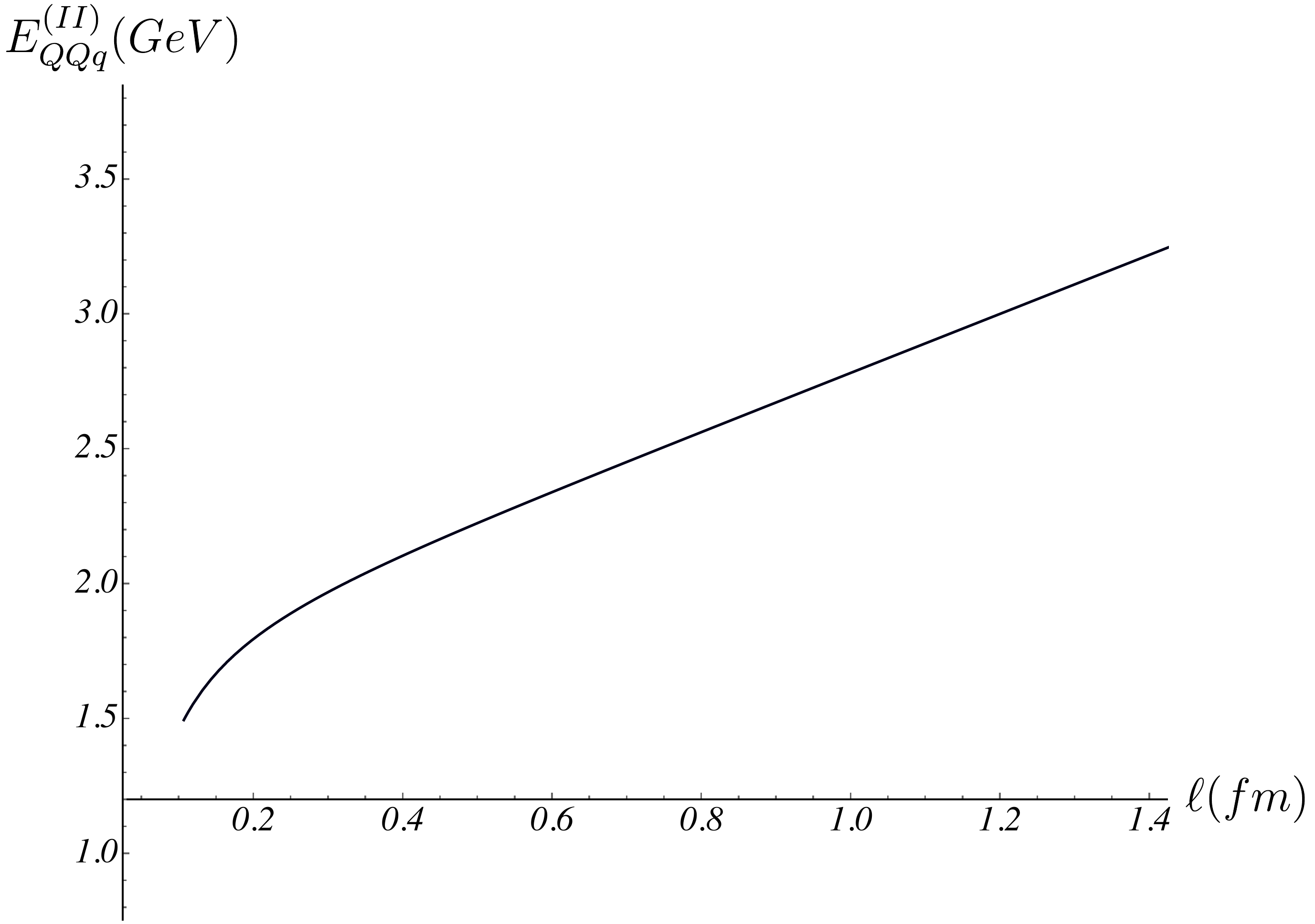}

\caption{{\small $\EII$ as a function of $\ell$ at $\k=-\frac{1}{4}\ep^{\frac{1}{4}}$.}}
\label{Ed}
\end{figure}

In Figure \ref{Ed}, we display our result for $\EII(\ell)$ to see how this function looks like.   
We use the $L$ set of the fit parameters. In this case, a simple estimate gives $l(\vv)\approx0.107\,\text{fm}$. But what happens for smaller $\ell\,$? Does there exist a completion for this configuration?

To answer these questions, we note that for $v=\vv$ configuration II is symmetric under the reflection with respect to the $yr$-plane. So it looks like that of Figure \ref{QQqs}. Because of this, it is natural to think of that symmetric configuration as a completion of configuration II at small separations. Clearly, this automatically gives the desired factorization \eqref{factor}. The picture then looks as follows. For very small $\ell$  the string configuration is symmetric. As $\ell$ reaches a critical value $\ell(\vv)$, string (3) starts to slide down to the right or to the left. This breaks the symmetry, and the configuration takes the form shown in Figure \ref{QQqd}.  

It is also instructive to look at a asymptotic behavior of $\EII(\ell)$ for small and large $\ell$. As discussed in the above paragraph, the small $\ell$ behavior is given by Eq.\eqref{E-small}. The behavior for large $\ell$ can be read from Eqs.\eqref{E2Qq-d} and \eqref{EQQb-large}. So, we get 

\begin{equation}\label{E2-large}
\EII(\ell)=\sigma\ell-2\g\sqrt{\s}\III
+2c+o(1)
\,,	
\end{equation}
with
\begin{equation}\label{cir2}
	\III=I_0+\oh\Bigl({\cal Q}(\vv)-{\cal Q}(q)-3\k\frac{\ep^{-2\vv}}{\sqrt{\vv}}-\n\frac{\ep^{\oh q}}{\sqrt{q}}\Bigr)
	\,.
\end{equation}
Here $\sigma$ is the same string tension as we found earlier for configuration I. This is the expected result if the linear term comes from a long string connecting a diquark $[Qq]$ with the remaining heavy quark $Q$. As we will see below, this is indeed the case.

\subsubsection{A look at a quark-diquark picture $Q[Qq]$}

Before going on with the connected string configurations, it is worth saying a few words regarding a heavy-light diquark $[Qq]$.\footnote{For a review of much of what is known about diquarks, see \cite{diquarks}.} In the current context, it is in the antitriplet representation of the color group $SU(3)$. 

If one thinks of a diquark $[Qq]$ as a configuration of two strings stretched between the vertex and these quarks, then it is natural to suppose that the binding energy $E_{\Qd}$ of a diquark-quark system is given by that of the remaining string (2). Using Eqs.\eqref{EQQb} and \eqref{EQQb-part}, one finds that 

\begin{equation}\label{EQd}
	E_{\Qd}=\g\sqrt{\frac{\s}{\lambda}}\biggl(\int_0^1 \frac{du}{u^2} 
\Bigl(\ep^{\lambda u^2} \Bigl[1 - u^4 \ep^{2\lambda(1 - u^2)}\Bigr]^{-\oh}
 - 1 - u^2\Bigr)
 +
 \int_{\sqrt{\frac{\vv}{\lambda}}}^1 \frac{du}{u^2} 
\ep^{\lambda u^2} \Bigl[1 - u^4 \ep^{2\lambda(1 - u^2)}\Bigr]^{-\oh}
\biggr)+c\,.
\end{equation}
It is easy to see from this formula that for large $\ell$ the binding energy is given by 

\begin{equation}\label{Ed-large}
E_{\Qd}(\ell)=\sigma\ell-\g\sqrt{\s}I_{\Qd}+c+o(1)
\,,	
\end{equation}
with 
\begin{equation}\label{CQd}
I_{\Qd}=I_0
+
\int^{\sqrt{\vv}}_0 du\, u^2 \ep^{2-u^2} \Bigl[1 - u^4 \ep^{2(1 - u^2)}\Bigr]^{-\oh}
-
\int^1_{\sqrt{\vv}}\frac{du}{u^2} \ep^{u^2} \Bigl[1 - u^4 \ep^{2(1 - u^2)}\Bigr]^{\oh}
\,.
\end{equation}
In this derivation, we have assumed that a relative distance between the diquark and quark is defined by $\ell$.\footnote{In other words, the center of mass of the diquark coincides with the heavy quark position. That is certainly true for $m_Q\gg m_q$.} The first term is linear in $\ell$, as expected from the string theory constructions of quark potentials. Its coefficient is universal and is given by the physical string tension.   

\begin{figure}[htbp]
\centering
\includegraphics[width=7cm]{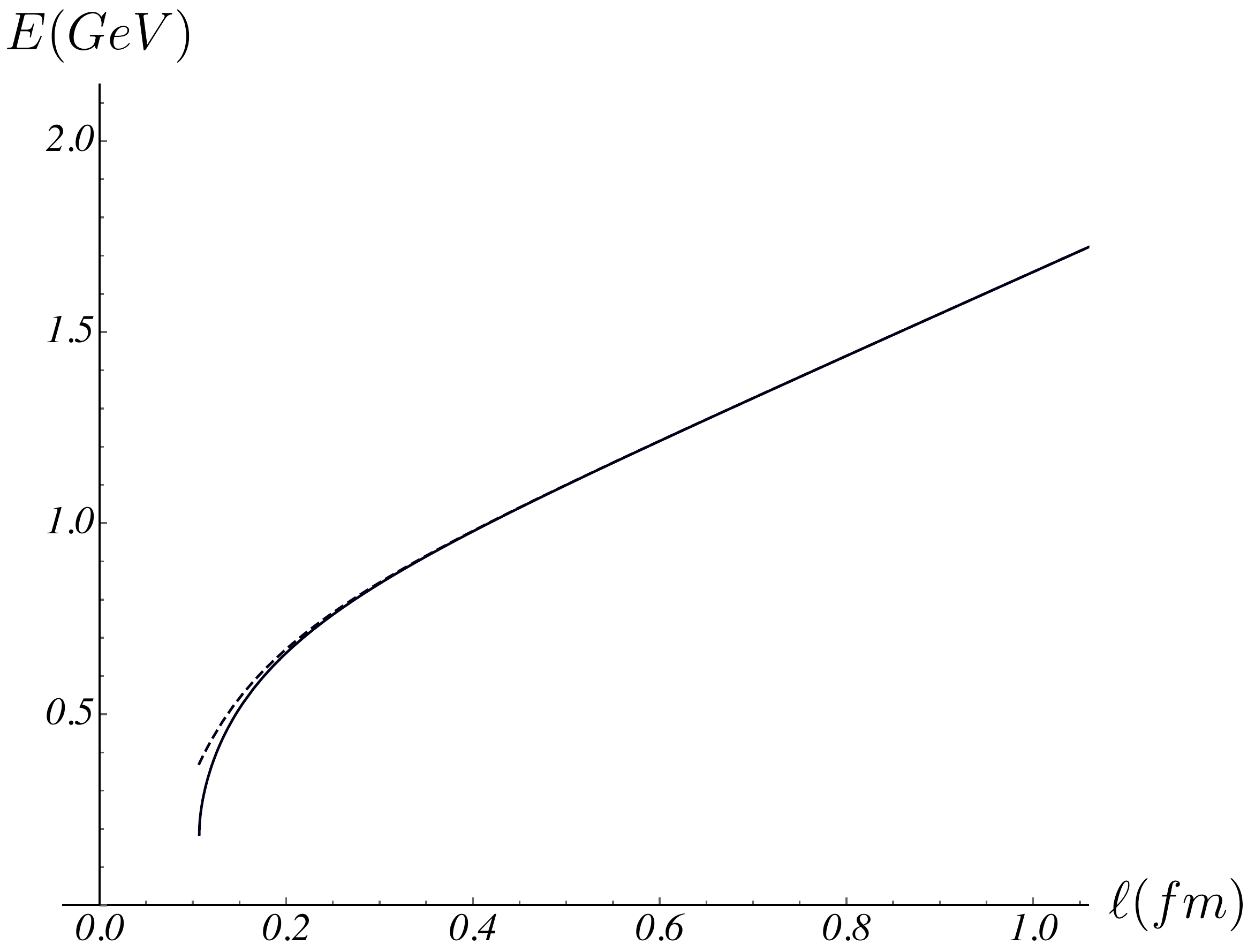}
\hspace{2 cm}
\includegraphics[width=7cm]{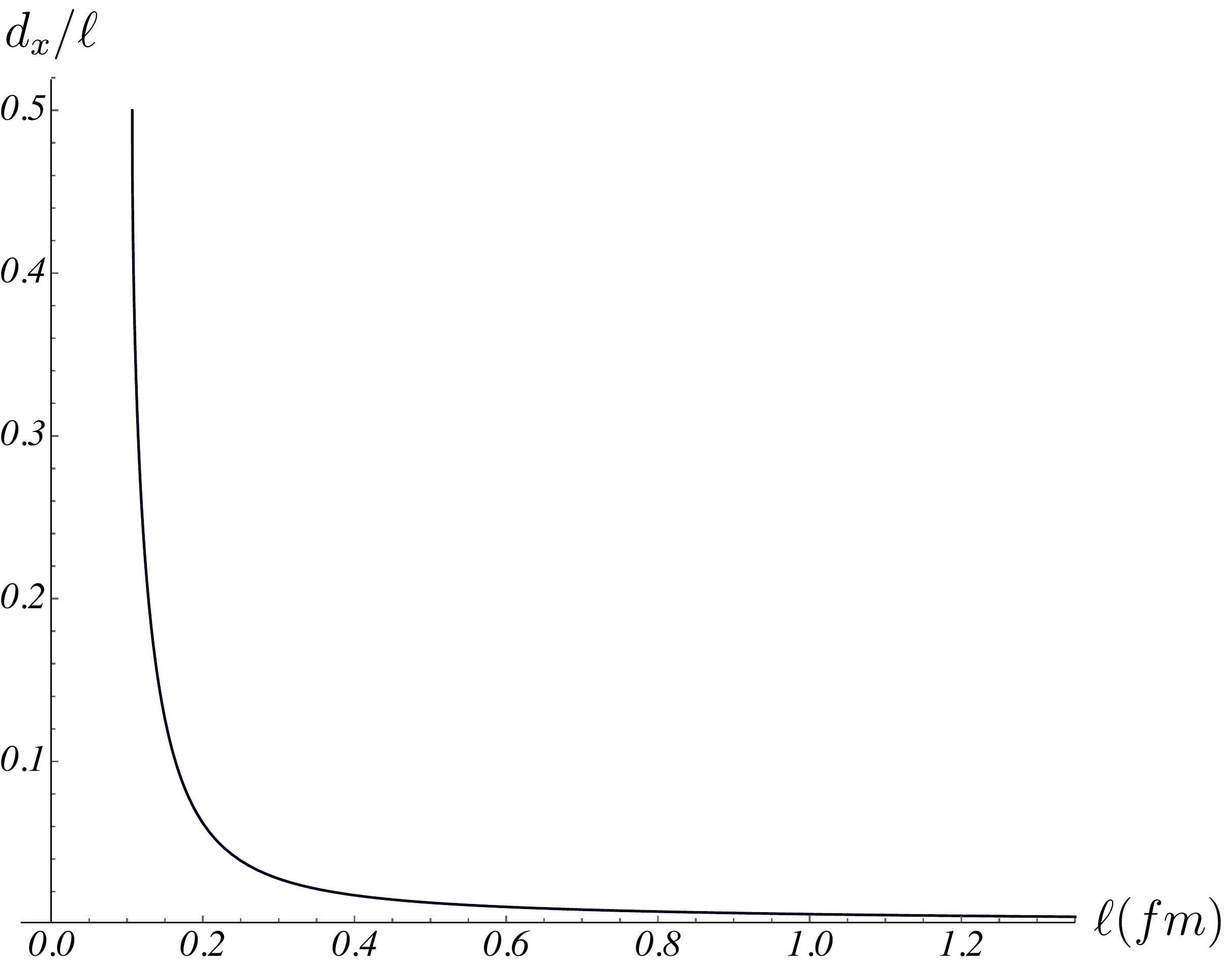}
\caption{{\small Left: Energies $E_{\Qd}$(solid curve) and $E_{\QQb}$ (dashed curve) vs $\ell$. Right: Ratio $d_x/\ell$ vs $\ell$. In both cases $\k=-\frac{1}{4}\ep^{\frac{1}{4}}$.}}
\label{E-Ld}
\end{figure}

It is of interest to see how strong the functional form of $E_{\Qd}$ deflects from that of $E_{\QQb}$. The result of such a comparison becomes more transparent if the normalization constant is set to $c=\g\sqrt{\s}(2I_0-I_{\Qd})$. In that case, the constant terms in the large $\ell$ expansions of $E_{\Qd}$ and $E_{\QQb}$ are equal to one another. For the parameter set $L$, with the redefined parameter $c$, the result is presented in the left panel of Figure \ref{E-Ld}. 
We see that the two forms are very similar to each other for $\ell\gtrsim 0.2\,\text{fm}$. This suggests that the quark-diquark picture of type $Q[Qq]$ is already valid for those distances. 

Assuming that the center of a light quark cloud is at $x=x_v$, we can see how far from the nearest heavy quark the center is. It follows from Figure \ref{QQqd} that a deviation in the $x$-direction is given by $d_x=x_v+\oh\ell$. This formula can be made explicit using \eqref{EQQb-part}. So

\begin{equation}\label{Ld}
d_x=\sqrt{\frac{\lambda}{\s}}
\int_0^{\sqrt{\frac{\vv}{\lambda}}} du\,u^2 \ep^{\lambda (1 - u^2)}
\Bigl[1-u^4\ep^{2\lambda (1 - u^2)}\Bigr]^{-\oh}
\,.
\end{equation}
The deviation defined this way has one important property: it depends on the separation between the heavy quarks. This follows from the parametric equation $\ell=\ell(\lambda)$.

One of the necessary conditions for the diquark approximation to be reliable is $1\gg d_x/\ell$. In Figure \ref{E-Ld}, we plot this ratio as a function of the heavy quark separation. As before, we use the $L$ set of the fit parameters. We see that the ratio is equal to $0.5$ at the minimal value of $\ell$, which is approximately $0.107\,\text{fm}$. Then it drops steeply to values below $0.1$ for $\ell\gtrsim 0.17\,\text{fm}$. In particular, it is about $0.081$ at $\ell=0.2\,\text{fm}$. This is another evidence for the quark-diquark picture at separation distances of over $0.2\,\text{fm}$. Finally, let us note that the cloud center is not exactly on top of the heavy quark, but it approaches this position as the quark separation is increased. 

\subsection{Which connected configuration is relevant?}

\begin{figure}[htbp]
\centering
\includegraphics[width=7.25cm]{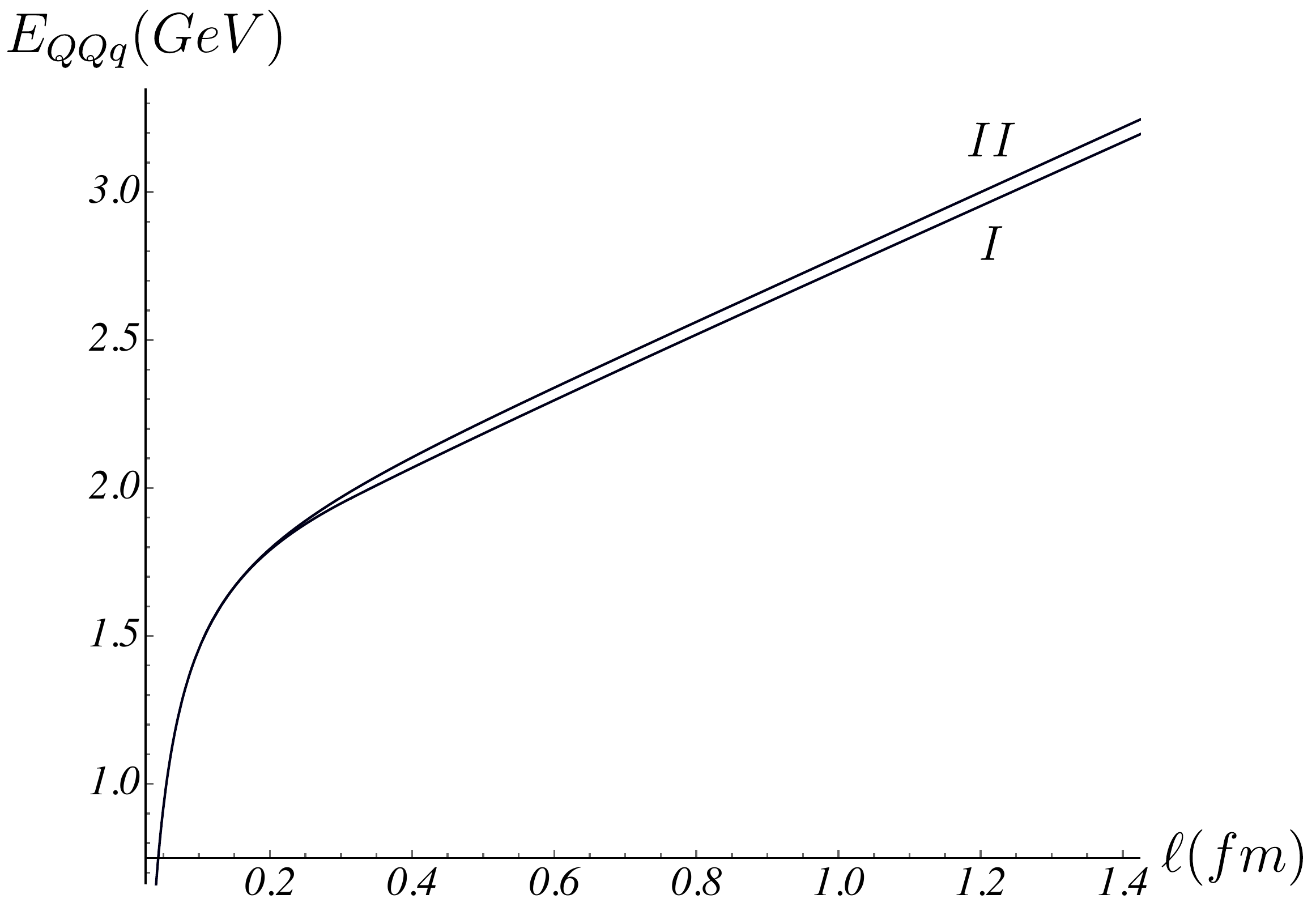}
\hspace{1.35cm}
\includegraphics[width=7cm]{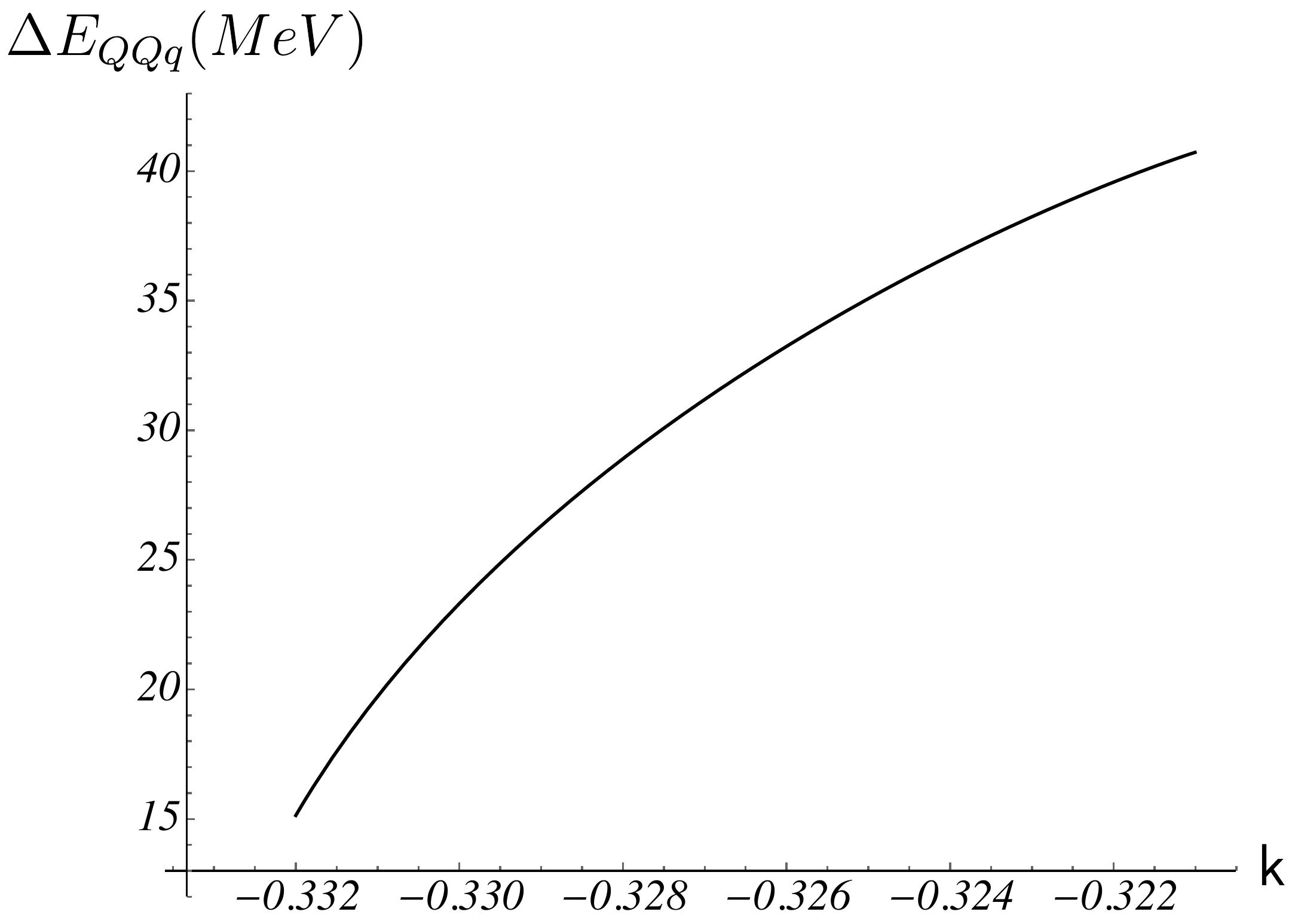}
\caption{{\small Left: The energies $\EI$ and $\EII$ at $\k=-\frac{1}{4}\ep^{\frac{1}{4}}$. Right: $\Delta E_{\QQq}$ as a function of $\k$.}}
\label{EIEII}
\end{figure}

A natural question to ask about the connected configurations is which configuration is energetically favored? To answer this question, we plot the energies $\EI$ and $\EII$ as functions of $\ell$ for the parameter set $L$. The result is presented in the left panel of Figure \ref{EIEII}. 
We see that the string configuration I is energetically favorable, but the gap is quite small. Such a gap can be analytically computed for large enough separations between the heavy quarks. From the asymptotic formulas \eqref{EQQq-large} and \eqref{E2-large}, one finds 

\begin{equation}\label{GapI}
\Delta E_{\QQq}=\g\sqrt{\s}
\biggl({\cal Q}(q)-{\cal Q}(\vv)
-2\int^1_{\sqrt{v_1}}\frac{du}{u^2}\ep^{u^2}\Bigl[1-u^4\ep^{2(1-u^2)}\Bigr]^{\frac{1}{2}}
+3\k\Bigl[\frac{\ep^{-2\vv}}{\sqrt{\vv}}-\frac{\ep^{-2v_1}}{\sqrt{v_1}}\Bigr]
+\n\Bigl[\frac{\ep^{\oh q}}{\sqrt{q}}-\frac{\ep^{\oh v_1}}{\sqrt{v_1}}\Bigr]
\biggr)
\,.
\end{equation}
For a non-zero solution $\vv$, which increases with increasing $\k$, the energy gap is shown in the right panel of Figure \ref{EIEII}.\footnote{This solution exists on the interval $-\frac{1}{3}< \k\leq-\frac{1}{4}\ep^{\frac{1}{4}}$.} Here we use the $L$ parameter set. The gap is increasing from $15\,\text{MeV}$ in the vicinity of $\k=-\frac{1}{3}$ to $41\,\text{MeV}$ at $\k=-\frac{1}{4}\ep^{\frac{1}{4}}$. 

Our conclusion is that in the model under consideration the quark-diquark picture of type $Q[Qq]$ does not appear explicitly at the level of the ground state. It turns out that the ground state is dominated by the most symmetric string configuration.
\subsection{A disconnected configuration and string breaking}

It is well-known that the presence of light quarks leads to string breaking and therefore to strong decays of hadrons. In the case of interest the dominant decay mode is into a baryon and a meson

\begin{equation}\label{decay}
QQq\rightarrow Qqq\,+\,Q\bar q
\,.	
\end{equation}

As usual, when studying this phenomenon in string models, one encounters disconnected string configurations. These are nothing else 
\begin{figure}[htbp]
\centering
\includegraphics[width=6.5cm]{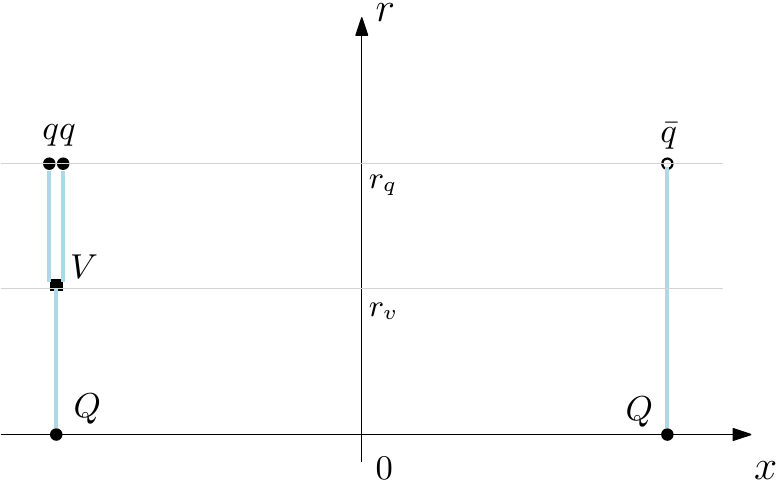}
\caption{{\small A disconnected static configuration. At zero baryon chemical potential, the light antiquark $\bar q$ is also at $r=\rq$.}}
\label{dis}
\end{figure}
but a stringy description of decay products. For our purposes in this paper, what we need is the static configuration shown in Figure \ref{dis}. It can be interpreted as a pair of non-interacting hadrons. One hadron is the heavy baryon $Qqq$ and another is the heavy-light meson $Q\bar q$. Each of those appeared separately in \cite{a-strb1}. Combining the corresponding expressions for $E_{\Qqq}$ and $E_{\Qqb}$, we find  

\begin{equation}\label{E-dis}
E_{\Qqq}+E_{\Qqb}=3\g\sqrt{\s}
\biggl({\cal Q}(q)-\frac{1}{3}{\cal Q}(\vv)+\n\frac{\ep^{\oh q}}{\sqrt{q}}
+\k\frac{\ep^{-2\vv}}{\sqrt{\vv}}\,
\biggr)+2c
\,.
\end{equation}
Here $q$ and $\vv$ are the solutions to Eqs.\eqref{fb-q} and \eqref{fbv2}, respectively. 

Having understood the string configurations relevant to the ground state of the $QQq$ system, it is straightforward to compute a string breaking distance. In doing so, an important fact, explained in \cite{drum}, will be that the phenomenon of string breaking can be studied by using a model Hamiltonian. In lattice QCD this Hamiltonian is usually extracted from a correlation matrix. For the problem at hand it is given by 

\begin{equation}\label{H}
H=
\begin{pmatrix}
\EI & g \\
g & E_{\Qqq}+E_{\Qqb} 
\end{pmatrix}
\,,
\end{equation}
where $\EI$ is the minimum energy of the bound system of three quarks and $E_{\Qqq}+E_{\Qqb}$ is the total energy of two non-interacting hadrons. The off-diagonal matrix element describes the mixing between these states. The important point emphasized in \cite{a-strb1} is that the diagonal elements of $H$ coincide with the energies of the corresponding string configurations. 

The string breaking distance emerges as a natural scale in characterizing this phenomenon. Like in \cite{drum, bulava}, we define it by simply equating the diagonal elements of $H$

\begin{equation}\label{lc}
\EI(\boldsymbol{\ell}_{\QQq})=E_{\Qqq}+E_{\Qqb} 
\,.	
\end{equation}
For large separations, where the phenomenon of string breaking is expected to occur, this equation simplifies drastically.\footnote{This is indeed the case, as can be seen from the estimates made below.} In this case $\EI$ becomes a linear function in $\ell$. So, using the expression \eqref{EQQq-large}, we get  

\begin{equation}\label{lc-Ex}
\boldsymbol{\ell}_{\QQq} =\frac{3}{\ep\sqrt{\s}}
\biggl(
{\cal Q}(q)-\frac{1}{3}{\cal Q}(\vv) +\n\frac{\ep^{\oh q}}{\sqrt{q}}+\k\frac{\ep^{-2\vv}}{\sqrt{\vv}}+\frac{2}{3}\II
\biggr)
\,.
\end{equation}
The result is scheme-independent as it should be. 

Now let us make a simple estimate of $\boldsymbol{\ell}_{\QQq}$. We take $\k=-\frac{1}{4}\ep^{\frac{1}{4}}$ simply because it leads to the exact solution $\vv=\frac{1}{12}$ of Eq.\eqref{fbv2}.\footnote{Another reason is that it is closest to phenomenologically motivated values of $\k$. We will return to this issue in Sec.IV.} For the parameter set $L$, we get 

\begin{equation}\label{lcL}
\boldsymbol{\ell}_{\QQq}=1.257\,\text{fm}
\,.	
\end{equation}
This value is very close (within $3\%$) to that obtained on the lattice for the meson decay mode $Q\bar Q\rightarrow Q\bar q+\bar Q q$, namely $\boldsymbol{\ell}_{\QQb}=1.220\,\text{fm}$ \cite{bulava}. The latter was used in \cite{a-strb1} to adjust the value of parameter $\n$. 

Because of the small deviation of these two values, it is of interest to ask the question: What happens if one uses another set of the parameters? To partially answer this question, we consider a phenomenologically motivated parameter set denoted in \cite{a-strb1} by $P$. It contains $\s=0.15\,\text{GeV}^2$, $\g=0.44$, and $\n=1.589$. So with these values

\begin{equation}\label{lcP}
\boldsymbol{\ell}_{\QQq}=1.073\,\text{fm}
\,,
\end{equation}
which is almost the same as the corresponding value $\boldsymbol{\ell}_{\QQb}=1.074\,\text{fm}$ for the meson mode \cite{a-strb1}.

Thus, this simple estimate suggests that the value of $\boldsymbol{\ell}_{\QQq}$ is very close or even equal to that of $\boldsymbol{\ell}_{\QQb}$. Hopefully, it will be possible eventually to check this prediction reliably by computer simulations. 

\subsection{The potential $V_0$}

\begin{figure}[htbp]
\centering
\includegraphics[width=7.7cm]{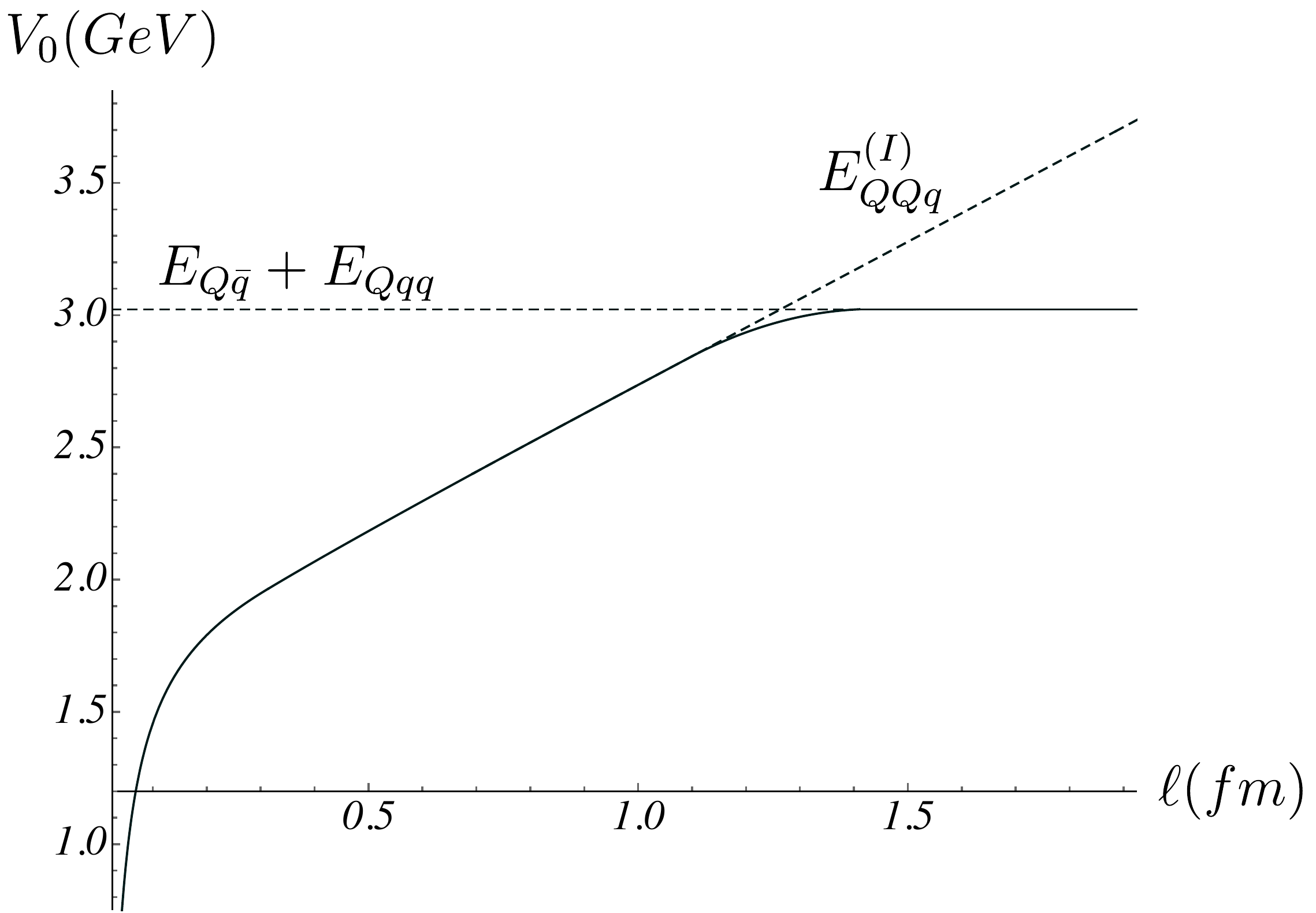}
\caption{{\small The potential of the $QQq$ system.}}
\label{V0}
\end{figure}

The ground state energy of the system that we call the potential $V_0$ is given by the smaller eigenvalue of the matrix Hamiltonian \eqref{H}. We can not find it without actually knowing the off-diagonal matrix element $g$. The latter can be in principle computed by means of lattice QCD, but how to do so within the effective string models remains unclear. In Figure \ref{V0}, we sketch the potential for the parameter set $L$ and $\k=-\frac{1}{4}\ep^{\frac{1}{4}}$. At  
any rate, it has the desired property of factorization at small $\ell$, as we discussed earlier in subsection A. At this stage, it is unclear whether one will be able to extract the string tension from the slope at intermediate distances. Certainly that depends on how large a transition area near $\boldsymbol{\ell}_{\QQq}$ is. 

\section{$E_{\QQ}$ and Lipkin rule}
\renewcommand{\theequation}{4.\arabic{equation}}
\setcounter{equation}{0}

The Lipkin rule is a simple way to relate the heavy quark-quark potential to the quark-antiquark potential by dividing the latter by two.  In the case of the Cornell potential $E_{\QQb}=-\frac{\alpha_{\QQb}}{\ell}+c+\sigma\ell$, this gives

\begin{equation}\label{Lipkin}
E_{\QQ}=-\oh\frac{\alpha_{\QQb}}{\ell}+c+\oh\sigma\ell
\,.	
\end{equation}
Only the coefficient of the Coulomb term comes from perturbative QCD (at least from the one-gluon exchange) that was the original motivation for this rule. The coefficient of the linear term related to the string tension still remains beyond the reach of perturbative QCD.

Having derived the expression for the heavy quark-quark potential in \cite{a-3q} and re-derived it in Sec.III again, we can now  attempt to elucidate the situation with Lipkin rule. In order for the rule to be valid to leading order in $\ell$, we require that $\k$ is a solution to $\alpha_{\QQ}(\k)=\oh\alpha_{\QQb}$, with the $\alpha$'s defined by Eqs.\eqref{E-small} and \eqref{alpha-QQb}. If so, then $\k\approx -0.102$ numerically. The constant term in \eqref{factor} is the same as required by the Lipkin rule. This happens automatically, without any additional assumptions. However, at next order, a simple estimate gives $\boldsymbol{\sigma}_{\QQ}= 0.819\,\sigma$ instead of the expected $0.5\,\sigma$. So, in the model we are considering the Lipkin rule breaks down at this order. 

The obvious question one might ask at this point is whether, instead of considering the Cornell-type potential for $E_{\QQb}$, one can find a relation similar to Lipkin rule but for the current model. The idea might be to reconsider the short distance behavior of $E_{\QQb}$. The point is that the Cornell potential has the same coefficient $\sigma$ in front of the linear terms for both short and long distance expansions, while the current model does not have this property. With the help of \eqref{alpha-QQb}, the coefficient $\boldsymbol{\sigma}_{\QQ}$ can easily be evaluated in terms of $\boldsymbol{\sigma}_{\QQb}$. At $\k=-0.102$ one finds numerically 

\begin{equation}\label{sigma-Lipkin2}
	\boldsymbol{\sigma}_{\QQ}=1.018\, \boldsymbol{\sigma}_{\QQb}
	\,.
\end{equation} 
This suggests that the relation we are looking for is 

\begin{equation}\label{Lip-new}
E_{\QQ}=-\oh\frac{\alpha_{\QQb}}{\ell}+c+\boldsymbol{\sigma}_{\QQb}\ell 
	\,.
\end{equation} 
Obviously, there is no simple relation between $E_{\QQ}$ and $E_{\QQb}$ which is similar to that of Lipkin. 

\begin{figure}[htbp]
\centering
\includegraphics[width=7.5cm]{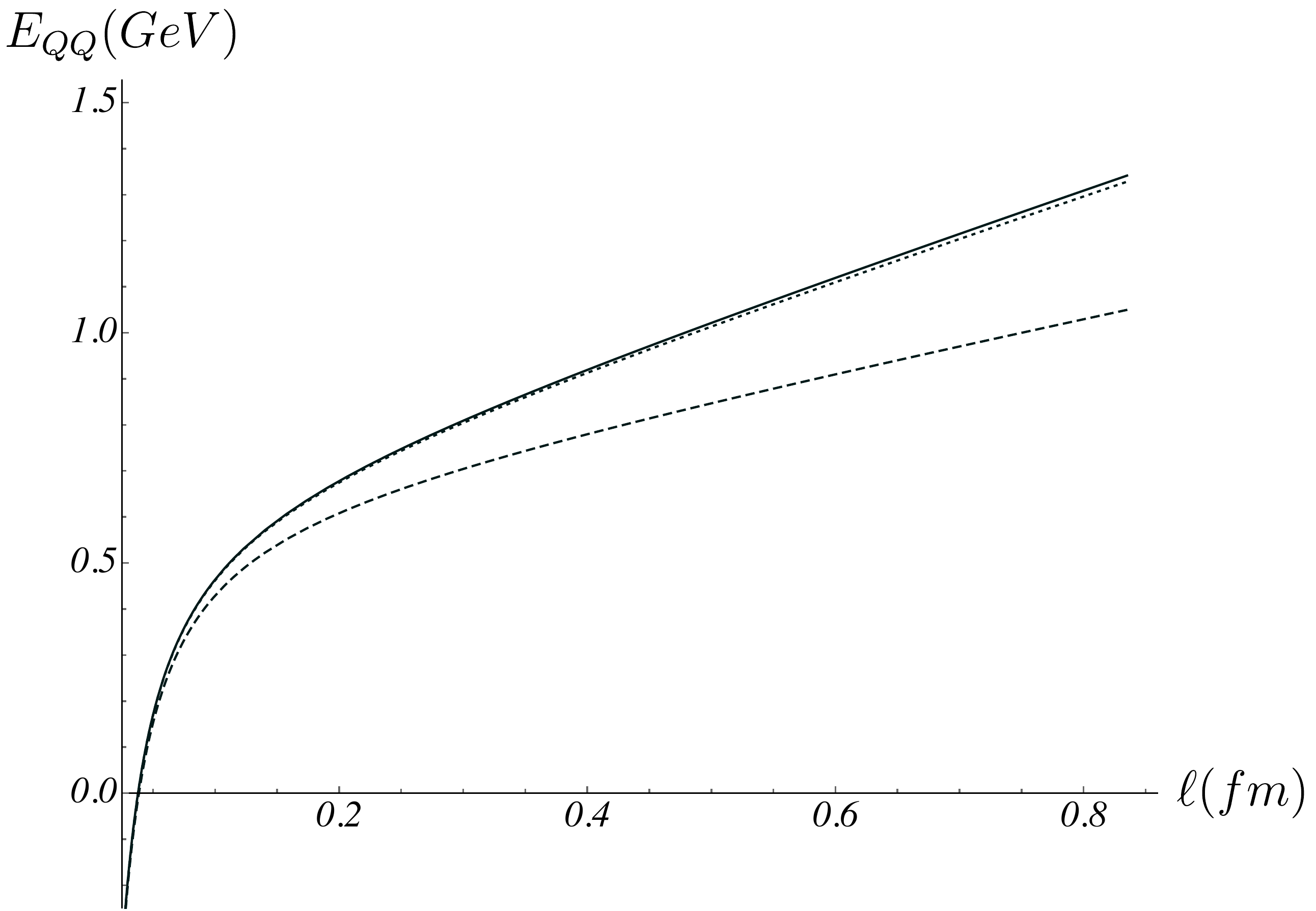}
\caption{{\small Various $E_{\QQ}$ vs $\ell$ plots. The curves correspond to \eqref{factor} (solid), \eqref{Lipkin} (dashed), and \eqref{Lip-new} (dotted).}}
\label{EQQ}
\end{figure}

To illustrate this, we plot the potential $E_{\QQ}$ together with the two approximations, which follow from Eqs.\eqref{Lipkin} and \eqref{Lip-new}. For the parameter set $L$, the results are presented in Figure \ref{EQQ}. We see that Lipkin rule is  
reasonably good only for separations less than $0.2\,\text{fm}$. On the other hand, the approximation \eqref{Lip-new} is accurate in the whole range, where the factorization holds, i.e., up to separations of order $0.8\,\text{fm}$ (see the right panel of Figure \ref{E2QqI}).\footnote{In general, the factorization and Lipkin rule are applicable in different ranges of $\ell$.} 

We conclude this discussion with a remark. The issue of the subleading linear term in the heavy quark-antiquark potential at small quark separations has been discussed in the literature for almost three decades.\footnote{For a brief introduction and references to the literature, see \cite{viz}.} One of the ideas is that an infinite  (or at least long enough to be considered as infinite) perturbative series can be parameterized by simply combining a Coulomb term with a subleading linear term. Our estimate suggests that the coefficients of the linear terms for $E_{\QQ}$ and $E_{\QQb}$ are almost or exactly equal to each other. In other words, in contrast to the Coulomb terms, the linear terms are universal in sense that there is no dependence on a specific representation of color $SU(3)$. At this point we would re-emphasize that neither $\boldsymbol{\sigma}_{\QQb}$ nor $\boldsymbol{\sigma}_{\QQ}$ is equal to the physical string tension $\sigma$.

\section{Concluding comments}
\renewcommand{\theequation}{5.\arabic{equation}}
\setcounter{equation}{0}

In this paper, we have advanced the use of the 5-dimensional effective string models for application to QCD. By doing so, we gained a better insight on the $QQq$ system. Apart from the expected factorization at small separations of heavy quarks, our model predicts that the string tension is exactly the same as in the $Q\bar Q$ and $QQQ$ systems and the string breaking distance is almost the same as that in the $Q\bar Q$ system. It is worth noting that we did not introduce any new parameter in addition to those for the $Q\bar Q$ and $QQQ$ potentials. The model we presented, however, is not exactly dual to QCD. It is an effective model which has its limitations and shortcomings, and therefore should be considered with caution. 

In general, the main question to be answered is What is the string dual to QCD? Is it a five-dimensional theory, as suggested in \cite{Pbook}, or a modification of AdS/CFT \cite{malda}? Our investigations of the $Q\bar Q$ and $QQQ$ potentials \cite{a-3q,az1} indicate that the five-dimensional description is quite accurate actually. On the other hand, it was shown in \cite{a-T2} that with the help of the geometry \eqref{metric} one may not correctly reproduce the $T^2$-term in the equation of state of pure gauge theory at finite temperature and therefore make contact with the lattice results. A possible way out is to assume the warping of five-dimensional internal space, i.e. the modification of AdS/CFT that leads to the desired result. If so, then a question arises concerning a Ramond-Ramond background and technicalities related with it. These issues are beyond the scope of this paper. Let us only note that one explanation of why the static force $\mathbf{f}_v$ pulls the baryon vertex towards the boundary is that Ramond-Ramond repulsion is stronger than graviton-dilaton attraction. Thus, from a ten-dimensional point of view $\mathbf{f}_v$ is a net force.\footnote{A possible alternative is a five brane of negative tension \cite{negT}. The standard objection to it is an instability due to small fluctuations. However, this is not so apparent in the $QQQ$ system where the baryon vertex is not an isolated brane but a part of the system.} 

In the real world, the number of colors is three. It is expected that the string description of baryons is valid for a large number of colors. The long-standing question is whether $N_c=3$ is large enough. We have no general arguments for or against it. In \cite{a-3q} it was shown that the string model describes the lattice data for the $QQQ$ potential pretty well. Is it also the case for the $QQq$ potential? Hopefully, it will be possible eventually to compute it reliably by numerical simulations and compare the result with our predictions.

Our construction of string configurations is general and independent of a specific background geometry. The reasons for choosing the particular geometry \eqref{metric} as an example are: (i) As noted in introduction, it provides the results consistent with the lattice QCD and phenomenology. (ii) It is simple and yields analytic results. (iii) It allows one to make predictions which may then be tested by means of other non-perturbative methods. In this regard it would be of interest to consider more realistic string models and answer the question: What else can string models say about the properties of the $QQq$ potential? 

We discussed the ground state of the $QQq$ system. From a group theoretical point of view, it corresponds to the representation of the cylindrical symmetry group with $\Lambda=0$ (the projection of the light quark angular momentum on the $QQ$ axis). By contrast, the representations with $\Lambda=\oh$ and $\Lambda=\frac{3}{2}$ were studied in lattice QCD \cite{bali}. This makes it impossible to make a direct comparison of the results. So, it seems important to us to develop string constructions for these excited energy levels as well. We expect that a heavy-light diquark $[Qq]$ may naturally occur in such constructions, as it does for excited baryons with large angular momenta \cite{wilczek}. 

\begin{acknowledgments}
 We are grateful to P. de Forcrand for numerous discussions on the quark potentials and J. Soto for correspondence. We also thank the Arnold Sommerfeld Center for Theoretical Physics for hospitality. This research is supported by Russian Science Foundation grant 20-12-00200 in association with Steklov Mathematical Institute.
\end{acknowledgments}

\appendix
\section{Some details about the quark-antiquark potential}
\renewcommand{\theequation}{A.\arabic{equation}}
\setcounter{equation}{0}

In this Appendix we give a brief summary of some results about the heavy quark-antiquark potential (the ground state energy of a static quark-antiquark pair) which are relevant for our discussion in Sec.III. For standard explanations, see \cite{az1,a-3q,a-strb1,a-strb2} whose conventions we follow, unless otherwise stated. 

\begin{figure}[htbp]
\centering
\includegraphics[width=6.5cm]{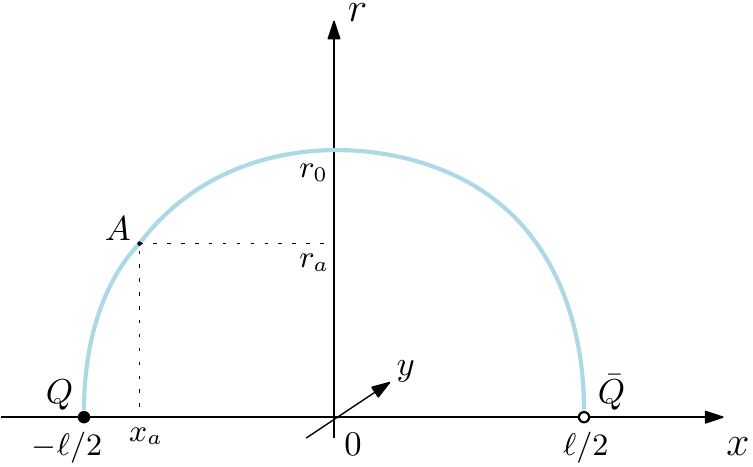}
\hspace{1.5cm}
\includegraphics[width=6.5cm]{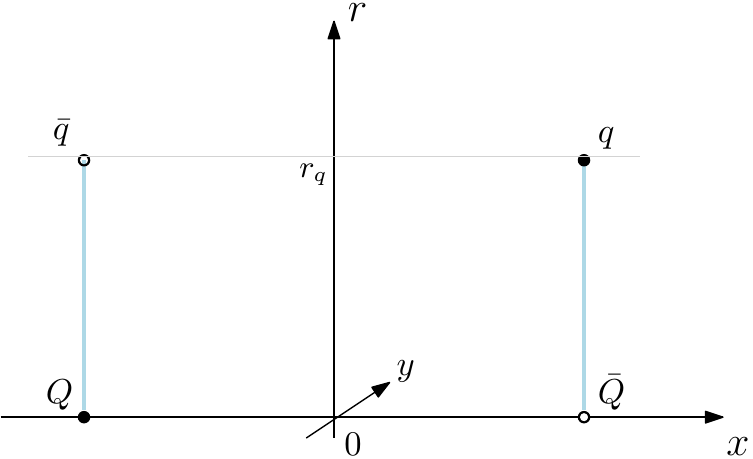}
\caption{{\small A selection of string configurations for the ground state energy of a static quark-antiquark pair. Here $r_0=r\vert_{x=0}$.}}
\label{QQb}
\end{figure}

First of all, let us consider the connected string configuration shown in Figure \ref{QQb}. It includes a string attached  
to the heavy quark sources on the boundary of space. For a Nambu-Goto string in the background geometry \eqref{metric}, the relation between the string energy and quark separation along the $x$-axis is written in parametric form

\begin{equation}\label{EQQb}
\ell= 2 \sqrt{\frac{\lambda}{\s}}
\int_0^1 du\,u^2 \ep^{\lambda (1 - u^2)}
\Bigl[1-u^4\ep^{2\lambda (1 - u^2)}\Bigr]^{-\oh}
\,,
\quad
E_{\QQb}=2\g\sqrt{\frac{\s}{\lambda}}
\int_0^1 \frac{du}{u^2} 
\Bigl(\ep^{\lambda u^2} \Bigl[1 - u^4 \ep^{2\lambda(1 - u^2)}\Bigr]^{-\oh}
 - 1 - u^2\Bigr)\,+2c\,,
\end{equation}
where $c$ is a normalization constant and $\lambda$ is a dimensionless parameter running from $0$ to $1$. It is expressed in terms of $\s$ and $r_0$ as $\lambda=\s r_0^2$. 

For a piece of the string between the points $Q$ and $A$, the corresponding formulas become 

\begin{equation}\label{EQQb-part}
\ell_{QA}=
 \sqrt{\frac{\lambda}{\s}}
\int_0^{\sqrt{\frac{a}{\lambda}}} du\,u^2 \ep^{\lambda (1 - u^2)}
\Bigl[1-u^4\ep^{2\lambda (1 - u^2)}\Bigr]^{-\oh}
\,,
\,\,
E_{QA}=\g\sqrt{\frac{\s}{\lambda}}
\biggl(\int_0^{\sqrt{\frac{a}{\lambda}}} \frac{du}{u^2} 
\Bigl(\ep^{\lambda u^2} \Bigl[1 - u^4 \ep^{2\lambda(1 - u^2)}\Bigr]^{-\oh}
 - 1 \Bigr)-\sqrt{\frac{\lambda}{a}}\biggr)+c\,.
\end{equation}
Here $\ell_{QA}=x_a+\frac{\ell}{2}$ and $a=\s r_a^2$. Note that the UV regulator used in the computations \cite{az1} leads to the lower bound for $a$.

The behavior of $E_{\QQb}$ for small $\ell$ is given by 

\begin{equation}\label{EQQb-small}
E_{\QQb}(\ell)=-\frac{\alpha_{\QQb}}{\ell}+2c+\boldsymbol{\sigma}_{\QQb}\ell +o(\ell)
\,,
\end{equation}
where 
\begin{equation}\label{alpha-QQb}
\alpha_{\QQb}=(2\pi)^3\Gamma^{-4}\bigl(\tfrac{1}{4}\bigr)\g
	\,,\qquad
	\boldsymbol{\sigma}_{\QQb}=\oh(2\pi)^{-2}\Gamma^{4}\bigl(\tfrac{1}{4}\bigr)\g\s
	\,.
\end{equation}
At the same time, for large $\ell$ it is 

\begin{equation}\label{EQQb-large}
E_{\QQb}(\ell)=\sigma\ell-2\g\sqrt{\s}\,I_0+2c+o(1)
\,,
\end{equation}
where
\begin{equation}\label{I0}
\sigma=\ep\g\s\,,
\qquad 
I_0=\int_0^1\frac{du}{u^2}\biggl(1+u^2-\ep^{u^2}\Bigl[1-u^4\ep^{2(1-u^2)}\Bigr]^{\frac{1}{2}}\biggr)
\,.
\end{equation}
Note that $I_0\approx 0.751$ and $\boldsymbol{\sigma}_{\QQb}/\sigma\approx 0.805$.

The disconnected configuration of the Figure provides a description of a pair of non-interacting heavy-light mesons. In this case, the total action has in addition to the Nambu-Goto actions of the fundamental strings contributions arising from the light quarks. It is thus $S=\sum_{i=1}^2 S_{\text{\tiny NG}}^{(i)}+S_{\text q}^{(i)}$. Varying the action with respect to $\rq$ gives the force balance equation \eqref{fb-q} for the light quarks. The energy of the configuration is given by 

\begin{equation}\label{2EQqb}
E_{\Qqb}+E_{\qQb}=2\g\sqrt{\s}\Bigl({\cal Q}(q)+\n\frac{\ep^{\oh q}}{\sqrt{q}}\,\Bigr)+2c
\,,
\end{equation}
with $q$ being a solution of Eq.\eqref{fb-q} in the interval $(0,1)$. Importantly, $c$ is the same normalization constant as that for the connected configuration. At zero baryon chemical potential, the meson's energies are equal one to another. 

By analogy with lattice gauge theory \cite{bulava}, the string breaking distance is defined by setting $E_{\QQb}(\boldsymbol{\ell}_{\QQb})=2E_{\Qqb}$. This equation simplifies for large quark separations, where $E_{\QQb}(\ell)$ is a linear function of $\ell$.\footnote{For the parameter values we use, this is true already for $\ell \gtrsim 0.5\,\text{fm}$, whereas the string breaking distance is about $1\,\text{fm}$.} If so, then the string breaking distance is  

\begin{equation}\label{lcm}
\boldsymbol{\ell}_{\QQb}=\frac{2}{\ep\sqrt{\s}}
\Bigl(
{\cal Q}(q)+\n\frac{\ep^{\oh q}}{\sqrt{q}}+I_0
\Bigr)
\,.
\end{equation}
It is important to note at this point that $\boldsymbol{\ell}_{\QQb}$ provides a scale for the decay process in which a heavy meson decays into a pair of heavy-light mesons: $Q\bar Q\rightarrow Q \bar q+\bar Q q$. 

The heavy quark-antiquark potential (the ground state energy of a static $Q\bar Q$ pair) asymptotically approaches $E_{\QQq}$ as $\ell$ tends to zero and $2E_{\Qqb}$ as $\ell$ tends to infinity. The transition between these two regimes occurs around $\ell=\boldsymbol{\ell}_{\QQb}$. The whole picture is similar to that of Figure \ref{V0}. 


\end{document}